\documentclass[prd, aps, floatfix, nofootinbib, superscriptaddress, twocolumn]{revtex4-2}
\usepackage{amsmath,amsfonts,amssymb,bm}
\usepackage{color}
\usepackage{dsfont}
\usepackage{dcolumn}
\usepackage{enumitem}
\usepackage{graphicx}
\usepackage{CJKutf8}
\usepackage{soul}
\usepackage{slashed}

\usepackage{color}

\usepackage{longtable}
\usepackage{multirow}
\usepackage{supertabular}
\usepackage{graphicx}

\usepackage[mathscr,scaled=1.15]{urwchancal}
\DeclareFontFamily{OT1}{pzc}{}
\DeclareFontShape{OT1}{pzc}{m}{it}%
{<-> s * [1.15] pzcmi7t}{}
\DeclareMathAlphabet{\mathpzc}{OT1}{pzc}{m}{it}

\definecolor{purple}{rgb}{0.5,0,0.5}
\definecolor{blue}{rgb}{0.0,0,0.9}
\definecolor{prdblue}{rgb}{0.133,0.118,0.498}
\usepackage[colorlinks=true, pdfstartview=FitV, linkcolor=prdblue, citecolor= prdblue, urlcolor=prdblue]{hyperref}

\newlength{\Wfockuud}

\def\s{\scriptscriptstyle}

\begin{document}
\begin{CJK}{UTF8}{song}
\title{$\,$\\[-7ex]\hspace*{\fill}{\normalsize{\sf\emph{Preprint no}. NJU-INP 058/22}}\\[1ex]
Hadron and light nucleus radii from electron scattering}

\author{Zhu-Fang Cui}
\email[]{phycui@nju.edu.cn}
\affiliation{School of Physics -- Nanjing University -- Nanjing Jiangsu 210093 -- China}
\affiliation{Institute for Nonperturbative Physics -- Nanjing University -- Nanjing Jiangsu 210093 -- China}
\author{Daniele Binosi}
\email{binosi@ectstar.eu}
\affiliation{European Centre for Theoretical Studies in Nuclear Physics
and Related Areas\\ Villa Tambosi -- Strada delle Tabarelle 286 -- I-38123 Villazzano (TN) -- Italy}
\author{\\Craig D.~Roberts}
\email[]{cdroberts@nju.edu.cn}
\affiliation{School of Physics -- Nanjing University -- Nanjing Jiangsu 210093 -- China}
\affiliation{Institute for Nonperturbative Physics -- Nanjing University -- Nanjing Jiangsu 210093 -- China}
\author{Sebastian M.~Schmidt}
\email[]{s.schmidt@hzdr.de}
\affiliation{Helmholtz-Zentrum Dresden-Rossendorf -- Dresden D-01314 -- Germany}
\affiliation{RWTH Aachen University -- III. Physikalisches Institut B -- Aachen D-52074 -- Germany}

\begin{abstract}
\leftline{\bf Abstract:}
\vspace*{-1.1em}

{\footnotesize
\hspace*{-1.55em}
Conceptually, radii are amongst the simplest Poincar\'e-invariant properties that can be associated with hadrons and light nuclei.  Accurate values of these quantities are necessary so that one may judge the character of putative solutions to the strong interaction problem within the Standard Model.  However, limiting their ability to serve in this role, recent measurements and new analyses of older data have revealed uncertainties and imprecisions in the radii of the proton, pion, kaon, and deuteron.  In the context of radius measurement using electron+hadron elastic scattering, the past decade has shown that reliable extraction requires minimisation
of bias associated with practitioner-dependent choices of data fitting functions.  Different answers to that challenge have been offered; and this perspective describes the statistical Schlessinger point method (SPM), in unifying applications to proton, pion, kaon, and deuteron radii.  Grounded in analytic function theory, independent of assumptions about underlying dynamics, free from practitioner-induced bias, and applicable in the same form to diverse systems and observables, the SPM returns an objective expression of the information contained in any data under consideration.  Its robust nature and versatility make it suitable for use in many branches of experiment and theory.}
\bigskip

\leftline{\bf Keywords:}
\vspace*{-1.15em}

{\footnotesize
\hspace*{-1.55em}
elastic electromagnetic form factors;
lepton scattering from hadrons and light nuclei;
emergence of mass;
muonic atoms;
proton;
Nambu-Goldstone modes -- pion and kaon;
deuteron;
strong interactions in the standard model of particle physics}

\end{abstract}

\date{2022 August 09}

\maketitle
\end{CJK}

\tableofcontents

\section{Importance of Radii}
\label{SecIntro}
``How big is it?" and ``How heavy is it?" are two questions posed within all fields of life and endeavour, with meanings tuned to the objects under consideration.  In physics, they typically address signature Poincar\'e-invariant characteristics of a particle or bound state.  We focus on systems governed by strong interactions within the Standard Model of particle physics (SM), which are supposed to be described by quantum chromodynamics (QCD) \cite{Marciano:1977su}; namely, hadrons -- baryons (proton, neutron, \emph{etc}.), mesons (pion, kaon, \emph{etc}.), and light nuclei.

The QCD Lagrangian is built upon the non-Abelian colour gauge group, SU$_{\rm c}(3)$, and expressed in terms of gluon and quark fields.  Gluons transform as the adjoint representation of SU$_{\rm c}(3)$ and quarks form the fundamental representation.  This means that gluons and quarks all carry a colour quantum number, a generalisation of the dynamical charge in quantum electrodynamics (QED).  Empirically, colour is \emph{confined}.  Consequently, whilst these gluons and quarks provide the basis for developing a rigorous QCD perturbation theory \cite{Brock:1993sz}, which is crucial in the analysis of high-energy processes, they are essentially different from the quanta of QED.  The gluons and quarks used to express the QCD Lagrangian are not the objects that propagate to be captured in detectors.  Instead, only composite bound-states, seeded by colour singlet combinations of QCD quanta, are directly measurable.

\begin{figure}[t]
\centerline{%
\includegraphics[clip, width=0.28\textwidth]{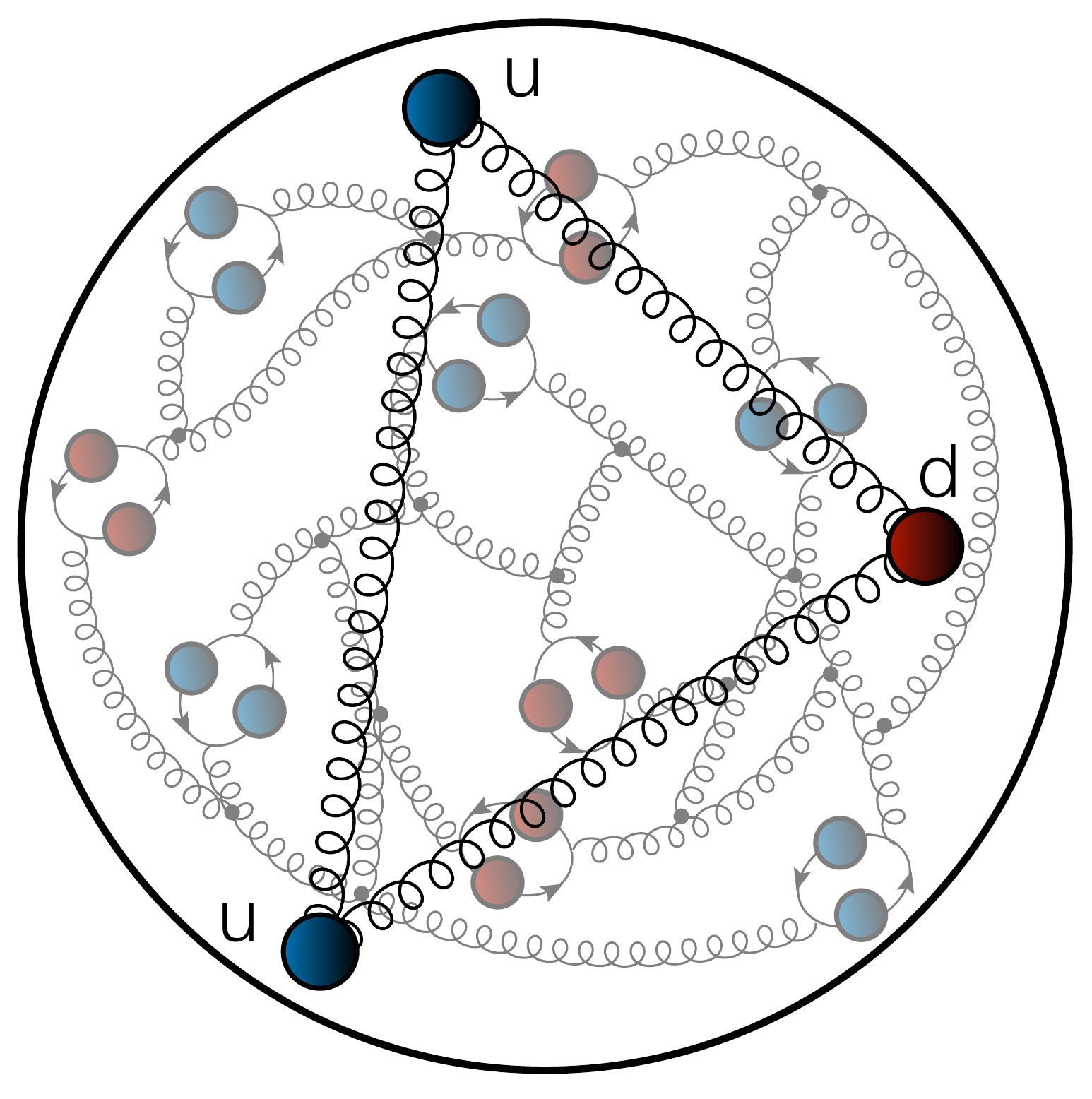}}
\caption{\label{Fproton}
In terms of QCD's Lagrangian quanta, the proton contains two valence up ($u$) quarks and one valence down ($d$) quark, the combination of which is a colour singlet.  Owing to the character of quantum field theory and the nature of strong interactions, the proton also contains infinitely many gluons and sea quarks, drawn here as ``springs'' and closed loops, respectively.  (The neutron looks similar, except that it is seeded by two $d$ quarks and one $u$ quark.)}
\end{figure}

The proton is a prime example, as highlighted by Fig.\,\ref{Fproton}.  Perceived in terms of QCD's Lagrangian degrees-of-freedom, it is a bound-state seeded by a colour singlet three-quark combination (two $u$ quarks and one $d$ quark) that is embedded in a complex medium produced by infinitely many gluons and quark+antiquark pairs that is also colour neutral overall \cite{Brodsky:2010xf}.  With so much stored within, it is natural to ask after the size of this highly nontrivial arrangement of quanta.

The natural scale for the size of atomic systems is the Angstrom, $\AA$.  Using natural units ($\hbar = 1 = c)$, this scale is set by $1/[m_e \alpha_{\rm em}]$, where $m_e\approx 0.511\,$MeV is the electron mass and $\alpha_{\rm em}\approx 1/137$ is the QED fine structure constant.  The quarks in QCD's Lagrangian have current masses, $m_{u,d}$, generated by the Higgs boson, that are on the same scale as $m_e$ \cite[Sec.\,59]{Zyla:2020zbs} and the electromagnetic interactions of quarks are also characterised by $\alpha_{\rm em}$.  However, if the size of the proton in Fig.\,\ref{Fproton} were measured on the $\AA$ scale, then there would be no notion of confinement because modern detectors have the capacity to directly image targets of this size \cite{PhysRevLett.110.213001}.

Of course, the proton is very much smaller than the hydrogen atom; in fact, it provides the nucleus of $^1\!$H, whose scale is measured in femtometres: 1\,fm = $10^{-5}\AA$.  Hence, all the activity in Fig.\,\ref{Fproton} is contained within a volume on the order of 1\,fm$^3$.  It is logical to ask the question: Why is the proton so small?

At this point, another issue arises.  The proton mass is known to a high level of precision, with a relative error of $\sim 10^{-8}$ \cite{Zyla:2020zbs}: $m_p = 0.9382720813(58)\,{\rm GeV} = 1/[0.210309\,{\rm fm}]$.  This value is commensurate with the associated natural length scale; but, surprisingly, it is more than 100-times larger than the sum of the typically quoted current-masses of the proton's valence quarks \cite[Sec.\,59]{Zyla:2020zbs}.  One must then also ask: Why is the proton so heavy?

The proton is small and the proton is heavy.  Within the SM, these features are expected to be explained by QCD.  However, this anticipation raises yet another issue.  Given that $m_p$ is so much larger than the quark current-masses generated by the Higgs boson, then the bulk of the proton mass should be understandable in the absence of the Higgs mechanism of mass generation; namely, within QCD defined without quark couplings to the Higgs.  The problem with this reasoning is that the associated Lagrangian does not possess any mass-scales: without quark current-masses, the chromodynamics Lagrangian is scale invariant.  This is readily verified and raises the key, unifying question \cite{Roberts:2016vyn}: How does QCD itself generate the single mass-scale that characterises the proton mass, the proton size -- 
which may be reckoned as
the characteristic confinement length, and, in fact, the natural scale for all nuclear physics?
Approved and planned experiments and facilities are targeting this puzzle, as described in Refs.\,\cite{Adams:2018pwt, Andrieux:2020, Horn:2016rip, Aguilar:2019teb, Brodsky:2020vco, Chen:2020ijn, Anderle:2021wcy, Arrington:2021biu, AbdulKhalek:2021gbh} and related theory analyses \cite{Du:2020bqj, Xu:2021mju, Sun:2021pyw}.

The question can be posed as the need to understand the phenomenon of emergent hadron mass (EHM).  To achieve that goal, a \emph{solution} of QCD will be necessary.  (It will only be sufficient if QCD is the true theory of strong interactions.)  Nonperturbative continuum methods and numerical simulations of lattice-regularised QCD are together providing plausible solution paths \cite{Roberts:2021nhw}, although nothing yet approaches the level of mathematical rigour specified in the formulation of the Yang-Mills Millennium Problem \cite[pp.\,129--152]{millennium:2006}. Notwithstanding the progress made, EHM, confinement, and possible connections between them, remain fundamental open questions.  Not least because confinement underpins proton stability: unlike all other bound states, no proton has ever been seen to decay in the roughly 14-billion years since the Big Bang.  In fact, the current lower limit on the proton lifetime is $1.6\times 10^{25}$ billion-years \cite{Super-Kamiokande:2016exg}.

Particle masses can be measured in a variety of ways, \emph{e.g}., considering long-lived systems with electric charge, like the proton, a Penning trap can serve as a reliable mass spectrometer \cite{Heisse:2019xnz}.  Many hadron masses are measured with precision.  However, masses are volume-averaged quantities, sensitive to the underlying dynamics but not a keen discriminator between theories.  For instance, in strong interaction phenomenology, many widely differing Hamiltonians can produce hadron spectra that are equivalent so far as modern experiments can tell; \emph{e.g}., consider and compare Refs.\,\cite{Brodsky:2014yha, Giannini:2015zia, Eichmann:2016yit, Qin:2019hgk, Qin:2020jig, Yin:2021uom}.

On the other hand, as highlighted above, radii reveal more about the underlying theory; but relative to masses, it is far more difficult to formulate and complete the reliable calculation of a radius within a given theory and particle radii are harder to measure.  Again using the proton as an example, discovery of its anomalous magnetic moment, $\kappa_p\approx 2$, in 1933 \cite{FrischStern1} showed that the proton is not an elementary Dirac fermion.  However, the first measurement of its electric radius was not made until the mid-1950s using electron+proton ($ep$) elastic scattering \cite[Table~II]{Hofstadter:1956qs}: $r_p=0.77\pm 0.10\,$fm.

Electron+hadron scattering remains an excellent radius measurement tool because QED and the electron (probe) are well understood and analyses involving isolated targets avoid the need to account for binding corrections as in bound-state calculations \cite{Eides:2000xc}.  However, it does have issues.  In order to extract a radius, one must measure the relevant form factor, $F(Q^2)$, with high precision on a material domain of squared-momentum transfer, $Q^2$, that reaches deeply toward $Q^2=0$ because for systems with $F(0)\neq 0$, the radius is obtained via
\begin{equation}
\label{EqRadius}
r^2 = \left. -6 \frac{d}{dQ^2} \ln F(Q^2)\right|_{Q^2=0}\,.
\end{equation}
Achieving the precision, reach to $Q^2 \approx 0$, and density of $Q^2$-coverage necessary for a reliable radius determination is very challenging.

Naturally, an object's mass and its radius are correlated; and any (putative) SM solution will deliver values for both.  Hence, precise knowledge of an array of hadron and light-nuclei masses and radii are necessary to set rigorous benchmarks for theory.  Herein, therefore, we describe a mathematical technique, applicable in the same form to a wide variety of measurements, and also calculations, that when applied to high-precision, dense data stretching to very low momentum transfers will return an objective result for the associated charge radius.  Commonly described as the statistical Schlessinger point method (SPM) \cite{PhysRev.167.1411, Schlessinger:1966zz, Tripolt:2016cya, Chen:2018nsg}, but equally well as a statistical multi-point Pad\'e approximant scheme,
the approach avoids any specific choice of fitting function in analysing data.

\begin{figure*}[!t]
	\includegraphics[width=0.99\linewidth]{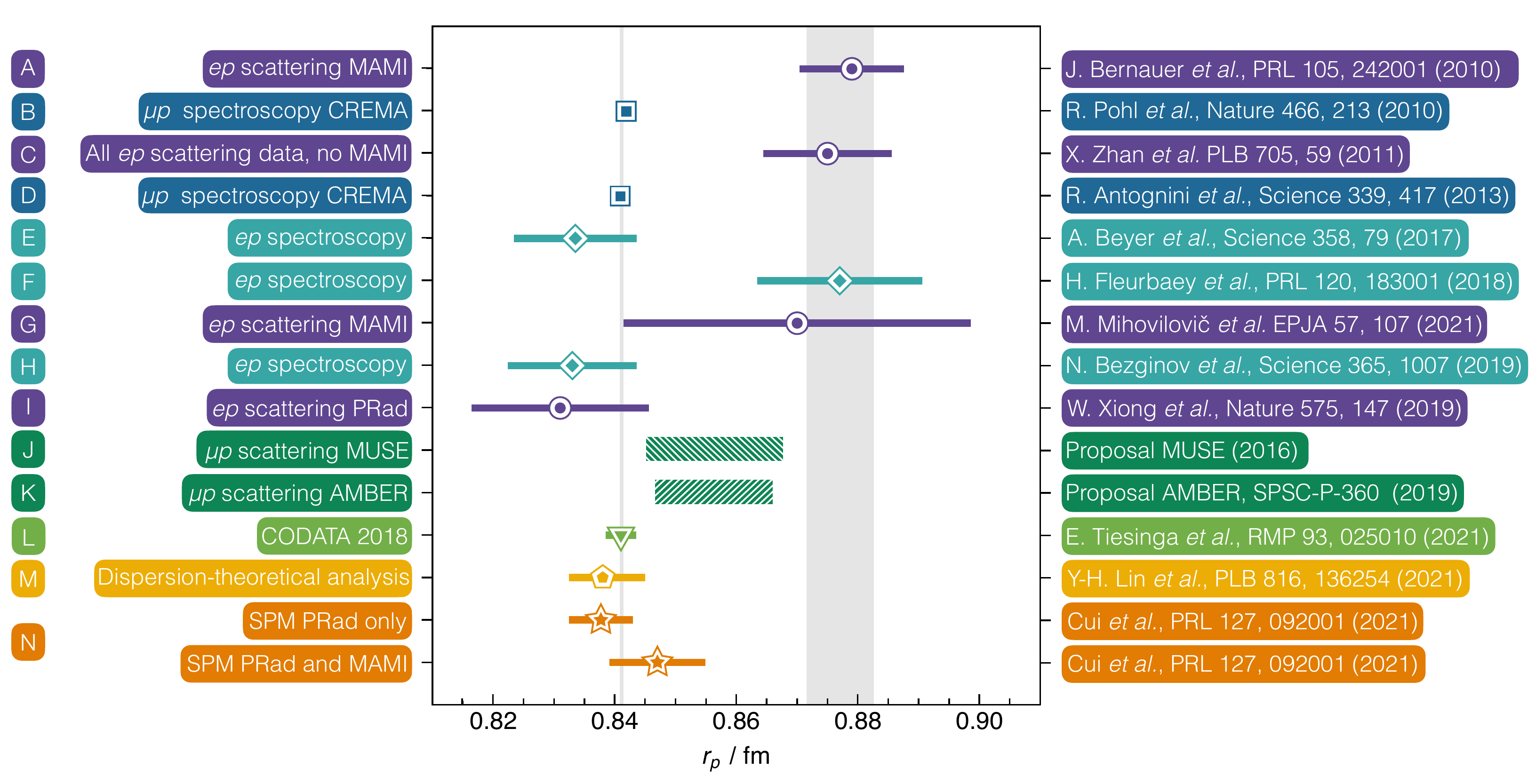}
	\caption{\label{fig:worldav}
Proton electric radius, $r_p$, extractions, various techniques:
[A] \cite{Bernauer:2010wm};
[B] \cite{Pohl:2010zza};
[C] \cite{Zhan:2011ji};
[D] \cite{Antognini:1900ns};
[E] \cite{Beyer:2017gug};
[F] \cite{Fleurbaey:2018fih};
[G] \cite{Mihovilovic:2019jiz};
[H] \cite{Bezginov1007};
[I] \cite{Xiong:2019umf}
[L] \cite{Tiesinga:2021myr};
[M] \cite{Lin:2021umk};
and [N] \cite{Cui:2021vgm}.
The green-hashed rectangles display the anticipated precision of two forthcoming muon+proton scattering experiments:
[J] \cite{Gilman:2013eiv, Cline:2021ehf};
and [K] \cite{Adams:2018pwt}.
The two vertical grey bands indicate the separate error-weighted averages of the small-radius and large-radius extractions.
}
\end{figure*}

\section{Proton Charge Radius -- Conflict}
In 2010, the Particle Data Group (PDG) listed the proton charge radius as \cite{Nakamura:2010zzi}: $r_p = 0.8768(79)\,$fm.  However, controversy was brewing following an extraction of the radius from measurements of the Lamb shift in muonic hydrogen ($\mu H$) \cite{Pohl:2010zza}: the result $r_p=0.84184(67)\,$fm is 5.0 standard deviations ($5.0\sigma$) smaller than the then accepted value.  As highlighted by Fig.\,\ref{fig:worldav}\,--\,Rows~[A\,--\,I], this disagreement stimulated many new experiments, with conflicting results; so, today, we are dealing with the ``proton radius puzzle'' \cite{Gao:2021sml}.

Although, the proton's mass can now be calculated with fair accuracy using modern theoretical tools \cite{Durr:2008zz, Eichmann:2016yit, Qin:2020rad}, the same is not true of $r_p$.  This deficiency has left room for many possible solutions of the puzzle to be offered, \emph{e.g}.\ \cite[Sec.\,5]{Carlson:2015jba}:
unknown QCD-related corrections may have been omitted in the $\mu H$ analysis, whose inclusion might restore agreement with the electron-based experiments that give a larger value;
new interaction(s) or particle(s) outside the SM may lead to a violation of universality between electron ($e$) and muon ($\mu$) electromagnetic interactions;
or some systematic error(s) has (have) hitherto been neglected in the analysis of $ep$ scattering.

Results from the most recent $ep$ scattering experiment were released in 2019 \cite[PRad]{Xiong:2019umf}:
\begin{align}
	r^\mathrm{PRad}_p/{\rm fm}=0.831\pm0.007_\mathrm{stat}\pm0.012_\mathrm{syst}\,.
	\label{PRad-res}
\end{align}
This was the first published $ep$ scattering analysis to obtain a radius in agreement with that extracted from $\mu H$ measurements: see Fig.\,\ref{fig:worldav}\,--\,Row~I.  Notably, the PRad experiment implemented some significant improvements over earlier efforts, \emph{inter alia}:
stretching to $Q^2 = 2.1\times 10^{-4}\,$GeV$^2$, the deepest reach yet achieved;
and simultaneously covering an extensive low-$Q^2$ domain, \emph{viz}.\ $2.1\times 10^{-4} \leq Q^2/{\rm GeV}^2\leq 6 \times 10^{-2}$.
Furthermore, since the charge radius is obtained using Eq.\,\eqref{EqRadius}, with $F=G_E^p$, the proton elastic electromagnetic form factor, the PRad analysis paid careful attention to the effect of data fitting function on the extracted charge radius.  The prejudice such choices may introduce has been highlighted
\cite{Kraus:2014qua, Lorenz:2014vha, Griffioen:2015hta, Higinbotham:2015rja, Hayward:2018qij, Zhou:2018bon, Alarcon:2018zbz, Higinbotham:2019jzd, Hammer:2019uab}.

PRad selected an optimal function form by using a bootstrap procedure applied to pseudodata produced with fluctuations mimicking the $Q^2$-binning and statistical uncertainty of their experimental setup.  No knowledge of actual PRad data was introduced.  This procedure renders the PRad extraction robust; but it also means that, ultimately, a specific function form was chosen \cite{Yan:2018bez}.

Of particular note is the conflict between the PRad result for $r_p$ and the values obtained in analyses of all other $ep$ scattering data, Fig.\,\ref{fig:worldav}\,--\,Rows~[A, C, G].  In this connection, the data described in \cite[A1]{Bernauer:2010wm} may be chosen as the benchmark, owing to their reach toward $Q^2=0$, high-precision, and sheer quantity.  A comparative analysis of the PRad and A1 data therefore provides an ideal means by which to illustrate the strengths of the SPM.

\section{Data Interpolation and Extrapolation}
\label{SecTheory}
Suppose one has $N$ pairs, ${\mathsf D} = \{(x_i,y_i=f(x_i)), i=1,\ldots,N$\}, being samples of some analytic function, $f(x)$, at a given set of enumerable points.  In applying the SPM, one constructs a continued-fraction interpolation:
\begin{equation}
{\mathpzc C}_N(x) = \frac{y_1}{1+\frac{a_1(x-x_1)}{{1+\frac{a_2(x-x_2)}{\vdots a_{N-1}(x-x_{N-1})}}}}\,,
\end{equation}
whose coefficients, $\{a_i|i=1,\ldots, N-1\}$, are computed recursively and guarantee ${\mathpzc C}_N(x_i) = f(x_i)$, $i=1\,\ldots,N$.  Mathematically, the method provides an accurate reconstruction of $f(x)$ within a radius of convergence determined by that one of the function's branch points which lies closest to the real-axis domain containing the data sample.  For instance, consider a monopole form factor sampled precisely at $N>0$ points, then using any one of those points, the SPM will truly reproduce the function.  The continued fraction interpolation encapsulates both local and global aspects of the underlying function.  Here, the global quality is crucial: it validates use of the constructed interpolation outside the domain of available samples and thus legitimises use of Eq.\,\eqref{EqRadius}, \emph{i.e}., evaluation of the curves' slope at the origin.

In physical cases, one is presented with $N$ data that are statistically distributed around some curve which must be accurately reconstructed.  So long as $N$ is large, then one can introduce a powerful statistical aspect to the SPM.  To wit, one selects at random $M < N$ points from the set $\mathsf{D}$.  In typical applications, $6\lesssim M \lesssim N/2$.  Then, in theory, one can build $C(N,M)=N!/[M!(N-M)!]$ -- a huge number -- of distinct interpolating functions.  In practice, the useable number is reduced somewhat by imposing physical constraints on their behaviour.
Herein, we consider electric and magnetic form factors at spacelike momenta.  Sampled on this domain, such functions must be continuously differentiable.  This is the restriction we implement.  It is a weak constraint, so the useable number of interpolants remains very large.

Using PRad data obtained at the $1.1\,$GeV electron beam energy as an example: $N=33$; so $M\in [6,17]$ may be chosen, giving ${\cal O}(10^6 - 10^9)$ possible interpolators; and one can productively work with the first $5\,000$ that are continuously differentiable monotonic functions on the entire range of $Q^2$ data.  Importantly, no other constraints need be applied; especially, no unity constraint on $G_E^p(Q^2=0)$.

Rather than present a dry mathematical account, we continue with this concrete descriptive illustration.  An extrapolation to $Q^2=0$ is defined by each interpolating function, using which $r_p$ can be calculated via Eq.\,\eqref{EqRadius}.  Then, for every $M$, a value of $r_p$ follows as the average of all results obtained from the $5\,000$ curves.

In completing the analysis, one must also account for the experimental errors in the given data set.  This can be done by using a statistical bootstrap process.  Namely, one generates, \emph{e.g}., $1\,000$ replicas of each set by substituting each datum by a new point, Gaussian distributed at random around a mean specified by the datum itself with variance equal to its error.  This leads to a probability distribution function, ${\mathpzc N}(\mu,\sigma)$, characterising the $r_p$ values that is, to an excellent approximation, normal, with average $r^{\s M}_p$ and standard deviation $\sigma=\sigma_{r}^{\s M}$.  (If a constant distribution is used instead, the mean is unchanged and the uncertainty is reduced.  The Gaussian distribution therefore provides a more conservative uncertainty estimate.)

Further, the fact that $M$ is deliberately varied leads to a second uncertainty source, $\sigma_{\s{\delta\!M}}$, that can be estimated by changing $M\to M'$, repeating the procedure just described for this new $M'$-value, and evaluating the standard deviation of the $r^{\s M}_p$ distribution for different $M$ values.

Combining these considerations, then for a given data set, the SPM result is
\begin{align}
	&r_p\pm\sigma_r;
	&r_p=\sum_{j=1}^{n_{\s M}}\frac{r^{\s{M_j}}_p}{n_{\s{M}}};&
&\sigma_r=\Bigg[\sum_{j=1}^{n_M}\frac{(\sigma^{\s{M_j}}_r)^2}{n^2_{\s{M}}}
+\sigma_{\s{\delta\!M}}^2\Bigg]^{\frac12}.
	\label{SPMrp}
\end{align}
For the PRad $1.1\,$GeV beam example, one computes radii for each of the values $\{M_j=5+j\,\vert\ j = 1,\dots,n_{\s{M}};\,n_{\s{M}} = 12\}$, so that for any given data set there are 60-million 
values of $r_p$, every one calculated from an independent interpolation.  In this case, $\sigma_{\s{\delta\!M}}\ll \sigma_{r}^{\s M_j}$ for all $j$s in the range specified above \cite[Supplement]{Cui:2021vgm}.

Before applying the SPM to any set of real data, one final issue must be addressed.  Namely, in ideal situations, experimental data are statistically distributed around that curve which properly represents the observable.  Since they do not lie on the curve, empirical data should not be directly interpolated.
This feature can be handled by \emph{smoothing} with a \emph{roughness penalty}, that may be implemented following the ALGOL procedure detailed in Ref.\,\cite{Reinsch:1967aa} and made concrete in Ref.\,\cite[Sec.\,3]{Cui:2021vgm}.  The approach is characterised by a rigorously defined optimal roughness penalty, $\epsilon$.  Its value is a self-consistent output of the smoothing procedure: $\epsilon =0$ means the data are untouched by smoothing, whereas $\epsilon =1$ returns a linear least-squares realignment of the data.  In the PRad example, $\epsilon \simeq 0$.

\begin{table}[t]
\caption{\label{TabSPMprocedure}
The SPM extraction of a radius from any high-quality $N$-element set of electron+hadron scattering data usually proceeds in these four steps.}
\begin{center}
\begin{enumerate}[label=(\roman*)]
\item \label{ItemLabel} Generate $1\,000$ replicas for the given experimental central values and uncertainties.
\item Smooth each replica using the associated optimal parameter, $\epsilon$;
\item For each number of input points $M$, chosen in the range $6\lesssim M \lesssim N/2$, determine the distribution of proton radii $r^{\s{M}}_p$, associated $\sigma^{\s{M}}_r$, and overall $\sigma_{\s{\delta\!M}}$;
\item Combine this material to calculate the final result for the radius and (statistical) uncertainty through Eq.\,\eqref{SPMrp}.
\end{enumerate}
\end{center}
\end{table}

Distilling the preceding remarks, the SPM extraction of a radius from any high-quality $N$-element set of electron+hadron scattering data usually proceeds as enumerated in Table~\ref{TabSPMprocedure}.

With any such method, it is fair to ask whether the scheme is robust, \emph{i.e}., whether it can reliably extract a known value in a diverse array of relevant cases.  For the proton radius case, that was checked using a wide variety of models employed over time by various teams to fit the world's $ep$ scattering data \cite{Borkowski:1975ume, Kelly:2004hm, Arrington:2003qk, Arrington:2006hm, Bernauer:2013tpr, Ye:2017gyb, Alarcon:2017ivh} and therefrom generating a proton electromagnetic form factor $G_E^p$ with a known value for the radius.  Using those models, Ref.\,\cite{Cui:2021vgm} generated replicas with the binning in $Q^2$ and errors of the A1 \cite{Bernauer:2010wm} and PRad \cite{Xiong:2019umf} data.
In all cases, irrespective of the $G_E^p$ model employed, the SPM was shown to return the radius value used to generate the pseudodata, with results being practically independent of $M$, the number of initial input points.
In Sec.\,\ref{SecDeuteron}, this standard validation procedure is detailed in connection with a proposal to extract the deuteron charge radius \cite{JlabDRad}.

\section{Proton Charge Radius -- Reconciliation}
\label{Recon}
With the method defined, one may ask what the SPM can reveal about the discrepancy between the $ep$ scattering results for $r_p$: Fig.\,\ref{fig:worldav} -- [A, C] vs.\ [I].  The first test is a comparison between the value of $r_p$ quoted by the PRad Collaboration [I] and the SPM.

There are $N=33$ points of PRad $1.1\,$GeV beam energy data.  Analysed as described in Sec.\,\ref{SecTheory}, one finds \cite{Cui:2021vgm}:
\begin{subequations}
\begin{equation}
r^{{\rm PRad}_{1.1}}_p/{\rm fm}=0.842\pm0.008_{\mathrm{stat}}\,.
\label{PRad11}
\end{equation}
With the $2.2\,$GeV beam energy, PRad collected $N=38$ data, using which the SPM yields  \cite{Cui:2021vgm}:
\begin{equation}
r^{{\rm PRad}_{2.2}}_p/{\rm fm}=0.824 \pm 0.003_{\mathrm{stat}}\,.
\end{equation}
\end{subequations}
Evidently, treated independently using the SPM, the $2.2\,$GeV data lead to a lower value of $r_p$ and a smaller uncertainty (one-third the size) than the $1.1\,$GeV set.

Combining the data from both beam energies \cite{Cui:2021vgm}:
\begin{equation}
	r^{{\mathrm{PRad}}}_p/{\rm fm}=0.838\pm0.005_{\mathrm{stat}}\,,
	\label{SPM-Prad}
\end{equation}
which is displayed in Fig.\,\ref{fig:worldav}\,--\,[N].  In this result, the smaller error determined in analysing the beam-energy $2.2\,$GeV data has driven the overall uncertainty down to roughly 50\% of that in Eq.\,\eqref{PRad11}.  These observations accord with those made by the PRad Collaboration \cite[Supplement\,--\,Fig.\,S16]{Xiong:2019umf}.  As highlighted by Fig.\,\ref{fig:worldav}\,--\,[I] \emph{cf}.\ [N], within mutual uncertainties, Eq.\,\eqref{SPM-Prad} reproduces the published PRad result.

A natural comparison for PRad data is provided by the measurements reported in \cite[A1]{Bernauer:2010wm}.   They encompass $1\,400$ cross-sections, recorded at beam energies $0.18$, $0.315$, $0.45$, $0.585$, $0.72$, $0.855\,$GeV, and stretch to $Q^2=3.8 \times 10^{-3}\,$GeV$^2$, which is deep but still roughly 20-times larger than the smallest PRad value.

Considering first the low-$Q^2$ region, consisting of $N=40$ data in the interval
$0.0038 \leq Q^2/{\rm GeV}^2 \leq 0.014$,
the SPM returns \cite{Cui:2021vgm}:
\begin{align}
	r^{{\mathrm{A1}-\mathrm{low}Q^2}}_p/{\rm fm} =0.856\pm0.014_{\mathrm{stat}}\,.
	\label{SPM-A1}
\end{align}
Expanding the selection to include all available A1 data, the SPM yields the same central value but a larger uncertainty:
$r^{{\mathrm{A1}}}_p/{\rm fm}=0.857\pm0.021_{\mathrm{stat}}$.
In this instance, $\sigma_{\s{\delta\!M}}\sim \sigma_{r}^{\s M_j}$.  Thus, extending the range up to $Q^2\sim 1\,$GeV$^2$ merely introduces a perceptible sensitivity to the number of data sampling points, $M$.
Now, whereas the original $r_p$ estimate from the A1 Collaboration is \cite{Bernauer:2010wm}: $r^{\mathrm{A1}-\mathrm{coll.}}_p/{\rm fm} =0.879\pm0.005_\mathrm{stat}\pm0.006_\mathrm{syst}$, the function-form independent SPM reanalysis of the A1 data, Eq.\,\eqref{SPM-A1}, reconciles the PRad and A1 values and shows both to be consistent with the $\mu H$ experiments.

Since they are now consistent, it is sensible to combine the PRad and A1 results [Eqs.\,\eqref{SPM-Prad} and \eqref{SPM-A1}]:
\begin{align}
	r^\mathrm{SPM}_p=0.847\pm 0.008_\mathrm{stat}\ [\mathrm{fm}].
\end{align}
This value is the lowest entry in Fig.\,\ref{fig:worldav}\,--\,[N].  Evidently, the SPM, which produces form-unbiased interpolations of data as the basis for a well-constrained extrapolation, reveals that there is no discrepancy between the proton radius obtained from $ep$ scattering and that inferred from
the Lamb shift in muonic hydrogen -- $r_p = 0.84136(39)\,$fm \cite{Pohl:2010zza, Antognini:1900ns}, %
the modern measurement of the $2S \rightarrow 4P$ transition-frequency in regular hydrogen -- $r_p = 0.8335(95)\,$fm \cite{Beyer:2017gug},
and the Lamb shift in atomic hydrogen -- $r_p = 0.833(10)\,$fm \cite{Bezginov1007}.
These values also, therefore, match the combination of the latest measurements of the $1S\rightarrow 3S$ and $1S\rightarrow 2S$ transition frequencies in atomic hydrogen -- $r_p=0.8482(38)\,$fm \cite{Grinin1061} and even the muonic deuterium determination $r_p=0.8356(20)\,$fm \cite{Pohl1:2016xoo}.
A recent assessment of proton radii extractions settles on the following value \cite{Zyla:2020zbs}:
\begin{equation}
\label{protonradius}
r_p/{\rm fm} = 0.8409 \pm 0.0004\,.
\end{equation}

This illustration shows that the proton radius is not a puzzle.  Instead, it highlights that a reliable extraction of $r_p$ from $ep$ scattering experiments is possible, so long as one works with precise data, densely packed at very-low-$Q^2$, and adopts an analysis scheme that eliminates systematic error introduced by the use of specific, limiting choices for the functions employed to interpolate and extrapolate the data.


\section{Proton Magnetic Radius}
The proton is a relativistic composite object, so its electromagnetic interactions involve two form factors, Dirac -- $F_1$ and Pauli -- $F_2$ \cite{Hofstadter:1956qs}, in terms of which the electric and magnetic form factors are
\begin{equation}
G_E = F_1 - \frac{Q^2}{4 m_p^2} F_2\,, \;
G_M = F_1 + F_2\,.
\end{equation}
Three-dimensional Fourier transforms of $G_{E,M}$ were long interpreted as measures of the spatial distributions of electric charge and magnetisation inside the proton \cite{Sachs:1962zzc}.  Today, their interpretation has changed \cite{Miller:2010nz}; but as emphasised above, the importance of measuring these key dynamical characteristics of the proton is undiminished.

Notably, for a structureless, noninteracting fermion, the Dirac equation entails $F_2\equiv 0 \Rightarrow G_E\equiv G_M$.  Perturbative corrections in QED only introduce small modifications for elementary fermions \cite{Aoyama:2012wj, Aoyama:2012wk}; hence, any material differences between $G_E$ and $G_M$ are marks of compositeness.  As remarked in Sec.\,\ref{SecIntro}, the first signal of this for the proton was the discovery of its anomalous magnetic moment $\kappa_p :=\mu_p-1:= F_2(0) \approx 2$ \cite{FrischStern1}.  Nevertheless, as recently as twenty years ago, available $ep$ scattering data were consistent with $\mu_p G_E (Q^2)/G_M(Q^2) = 1$.  Only with the operation of an electron accelerator combining high energy, luminosity, and beam polarisation was it revealed that $\mu_p G_E(Q^2)/G_M(Q^2) \neq 1$ \cite{Jones:1999rz}.
It follows that proton electric and magnetic radii must be different; unless some mechanism leads dynamically to
\begin{equation}
\label{EqFoldy}
-\frac{F_2^\prime (0)}{\kappa_p} = -F_1^\prime(0) + \frac{\mu_p}{4 m_p^2} \,.
\end{equation}

Eq.\,\eqref{EqFoldy} is curious because if true, then the proton's Pauli and Dirac radii are not independent observables.  Instead, their difference is a positive number fixed by the proton Foldy term \cite{Foldy:1958zz}.
No symmetry constraints entail Eq.\,\eqref{EqFoldy}.  Thus, the result may only emerge as a consequence of SM dynamics; but the precision of today's theory is insufficient to answer the question of whether it does, or not.

\begin{figure}[t!]
\includegraphics[width=0.999\linewidth]{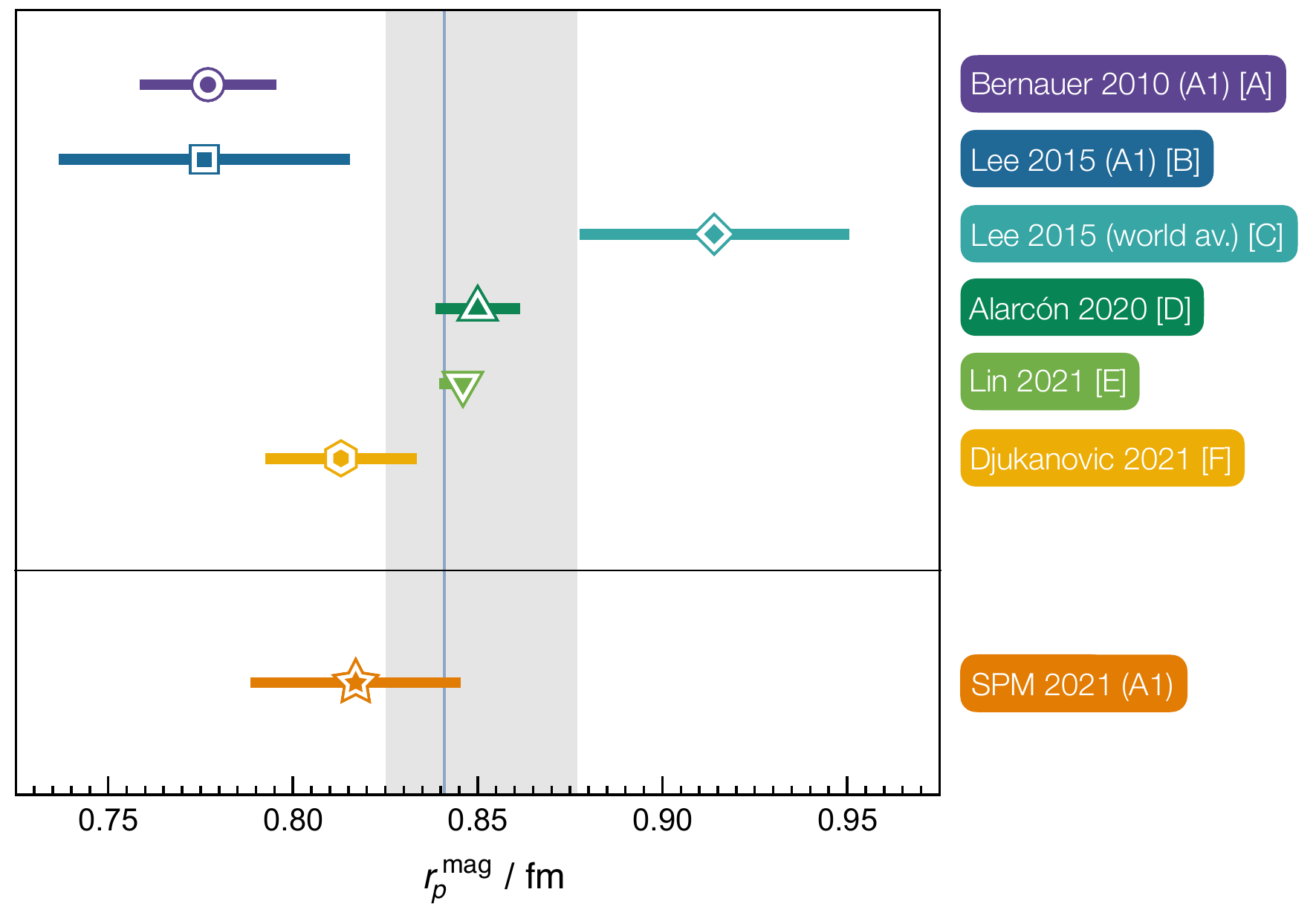}
	\caption{\label{FigMagRad}
{\it Upper frame}.  Proton magnetic radius extractions, various techniques:
[A] \cite{Bernauer:2010wm, A1:2013fsc} -- A1 Collaboration;
[B] \cite{Lee:2015jqa} -- A1 data;
[C] \cite{Lee:2015jqa} -- world average omitting A1 data;
[D] \cite{Alarcon:2020kcz} -- dispersion theory;
[E] \cite{Lin:2021umz} -- dispersion theory;
[F] \cite{Djukanovic:2021cgp} -- lattice QCD.
Light-grey band -- Eq.\,\eqref{BCaverage}; within mutual uncertainties, this value agrees with $r_p$ in Eq.\,\eqref{protonradius}, which is indicated by the thin vertical blue band.
{\it Lower frame}. SPM result in Eq.\,\eqref{SPMA1result}, obtained from A1 data \cite{Cui:2021skn}.}
\end{figure}

As emphasised by Fig.\,\ref{FigMagRad}, the empirical situation is also unclear.  The most precise data suitable for extraction of $r_p^{\rm mag}$ were obtained by the A1 Collaboration \cite{Bernauer:2010wm, A1:2013fsc}.  However, entries [B, C] in Fig.\,\ref{FigMagRad} highlight that when using precisely the same methods to analyse the world's data, markedly different values for $r_p^{\rm mag}$ can be obtained, depending on whether the A1 set is included or excluded.  Facing this conflict, the PDG \cite{Zyla:2020zbs} quotes an average of [B, C] in Fig.\,\ref{FigMagRad}:
\begin{equation}
\label{BCaverage}
r_p^{\rm mag}/{\rm fm} = 0.851 \pm 0.026\,;
\end{equation}
and given this controversy, it is reasonable to ask what the statistical SPM can objectively reveal about the proton magnetic radius from extant data \cite{Cui:2021skn}.

A1 data relate simultaneously to proton electric and magnetic radii.  The SPM result for the electric radius is discussed in connection with Eq.\,\eqref{SPM-A1}.  Regarding $r_p^{\rm mag}$, Ref.\,\cite[Supplement]{A1:2013fsc} lists $77$ points obtained via a Rosen\-bluth separation \cite{Rosenbluth:1950yq} of the $ep$ cross-sections into values for $G_{E,M}$ on $0.015 \leq Q^2/{\rm GeV}^2 \leq 0.55$.  The nine lowest-$Q^2$ $G_M$ data are widely scattered.  This is an issue with the Rosenbluth technique when applied to $G_M$ at low-$Q^2$.  Included in a SPM analysis, these nine points add noise without affecting the central value.  Thus, Ref.\,\cite{Cui:2021skn} reported a study based on the $N=68$ $G_M$ points on ${\cal D}_{A1} = \{Q^2\,|\,0.027 \leq Q^2/{\rm GeV}^2 \leq 0.55$\}.

The SPM analysis of $r_P^{\rm mag}$ follows the pattern employed for $r_p$.  In this case: $N=68$ and $M\in [7,19]$.  Consequently, there are O$(10^9 - 10^{15})$ potential interpolators; and steps (i)--(iv) in Table~\ref{TabSPMprocedure} are implemented within this space \cite{Cui:2021skn}.  As with $r_p$, the optimal smoothing parameter $\epsilon \simeq 0$ in all $r_p^{\rm mag}$-related cases.

An extensive set of SPM robustness tests in connection with A1 $G_M^p$ data revealed
\cite{Cui:2021skn} that for $M \in {\cal O}_M=\{ M_{ji} = 3 + 4 j + i\,\vert\, j=1,2,3,i=1,2,3\}$ one does not recover a satisfactory Gaussian distribution of radii.  (This contrasts with the $r_p$ case, where all values of $M$ are associated with true normal distributions.)  Consequently, only
$M\in {\cal S}_M=\{M_j=3+4j\,\vert\ j = 1,2,3,4\}$ were employed, leading to $20$--million results for $r_p^{\rm mag}$, each computed from an independent interpolation; hence,
\begin{equation}
\label{SPMA1result}
r_p^{\rm mag\;A1 - SPM}/{\rm fm} = 0.817 \pm 0.027_{\rm stat}\,.
\end{equation}
As apparent in Fig.\,\ref{FigMagRad}, within mutual uncertainties, this result agrees with that reported in Ref.\,\cite{Bernauer:2010wm}, although the central value is 5.1\% larger.  In comparison, the analogous SPM extraction of $r_p$ from A1 data, Eq.\,\eqref{SPM-A1}, returns a central result that is 2.6\% lower than originally reported \cite{Bernauer:2010wm}.

Now write
\begin{equation}
\label{EqFoldyApprox}
F_1^\prime(0) = {\mathpzc d}_1  \frac{\mu_p}{4 m_p^2}\,,  \;
\frac{1}{\kappa_p}F_2^\prime(0)  = [{\mathpzc d}_1-1-\delta_P]  \frac{\mu_p}{4 m_p^2}\,;
\end{equation}
Eq.\,\eqref{EqFoldy} is recovered with $\delta_P=0$.
Rigorous contact may be made between quantum field theory and quantum mechanics when one views the Poincar\'e-invariant Dirac and Pauli form factors from a light-front perspective \cite{Brodsky:1997de, Brodsky:2022fqy}.  Then \cite{Miller:2010nz}, $F_1^\prime(0)$ relates to the light-front transverse mean-square proton electric radius, $b_E^\perp$, and $F_2^\prime(0)$ to the analogous magnetic radius, $b_M^\perp$.  From this standpoint, Eq.\,\eqref{EqFoldyApprox} states that $b_M^\perp>b_E^\perp$ so long as $\delta_P>-1$.

\begin{figure}[t!]

	\includegraphics[width=0.725\linewidth]{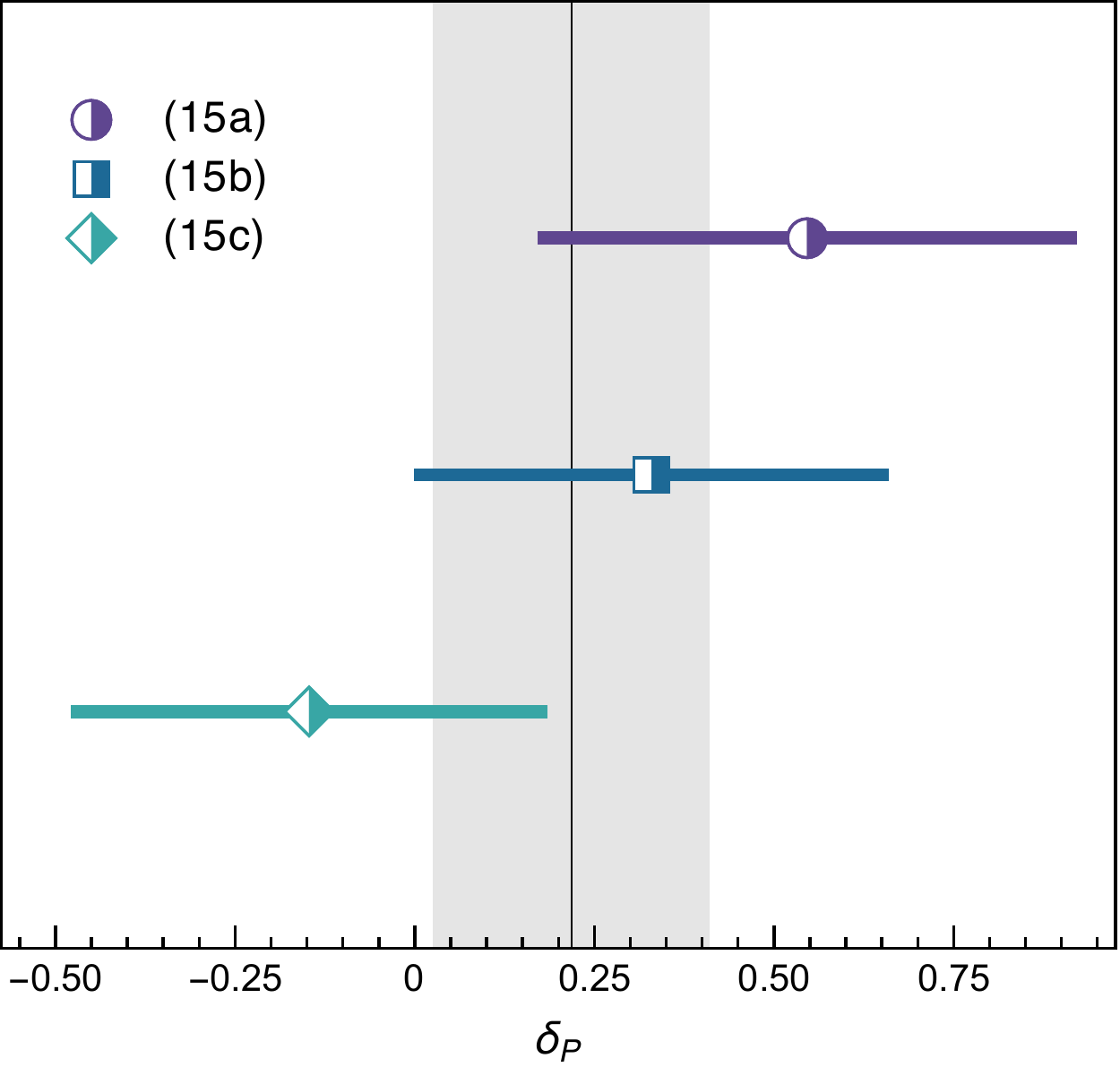}
\caption{\label{FdeltaP}
Experiment-based results for deviations ($\delta_P \neq 0$) from Eq.\,\eqref{EqFoldy} as listed in Eqs.\,\eqref{EqDelta1}.  The vertical grey line and associated band is the uncertainty-weighted average.
%
}
\end{figure}

Using the radii discussed above, the pairings
Eqs.\,(\ref{SPM-A1},\ref{SPMA1result}),
Eqs.\,(\ref{protonradius},\ref{SPMA1result}),
Eqs.\,(\ref{protonradius},\ref{BCaverage}) yield the results drawn in Fig.\,\ref{FdeltaP}:
\begin{subequations}
\label{EqDelta1}
\begin{align}
\delta_P^{{\rm A1}_{\rm SPM}} &= \phantom{-}0.546 \pm 0.366\,, \\
\delta_P^{\rm {\rm A1}_{\rm SPM} + PDG} &= \phantom{-}0.329 \pm 0.321 \,,\\
\delta_P^{\rm PDG } &= -0.147 \pm 0.322 \,.
\end{align}
\end{subequations}
The uncertainty-weighted mean of these values is $\delta_P=0.218 \pm 0.193$.
%
At low-$Q^2$, $[1-\mu_p G_E(Q^2)/G_M(Q^2)] \propto\delta_P$; so, improved accuracy might be attained through new high-precision low-$Q^2$ polarisation-transfer measurements of $\mu_p G_E(Q^2)/G_M(Q^2)$ \cite{Miller:2007kt, Cui:2021skn}.

Regarding Fig.\,\ref{FdeltaP}, it is clear that analyses of existing data are consistent with Eq.\,\eqref{EqFoldy} at a 26\% confidence level, \emph{i.e}., with this probability, the proton's Dirac and Pauli radii are not truly independent observables; so, Eq.\,\eqref{EqFoldy} is established as a useful approximation.  Moreover, the analyses reveal that the proton's Pauli radius is very probably larger than its Dirac radius.  These conclusions place valuable constraints on pictures of proton structure \cite{Cui:2021skn}.  This is important because there are many possible contributions to $F_2(Q^2\simeq 0)$, including \cite{Punjabi:2015bba, Cloet:2013jya, Eichmann:2016yit, Brodsky:2020vco, Barabanov:2020jvn, Giannini:2015zia, Sufian:2016hwn, Xu:2019ilh, Mondal:2019jdg, Cui:2020rmu, Xu:2021mju}: quark anomalous magnetic moments, orbital angular momentum within the proton, meson cloud contributions, \emph{etc}.; and there are related influences on $F_1(Q^2\simeq 0)$. An insightful understanding of the accuracy of Eq.\,\eqref{EqFoldy} will require an explanation of how these factors combine to yield a simple (approximate) algebraic outcome; and, tied to that, experiments and analyses that supply a precise result for $r_p^{\rm mag}$. 

\begin{figure}[t]
\centerline{\includegraphics[width=0.28\textwidth]{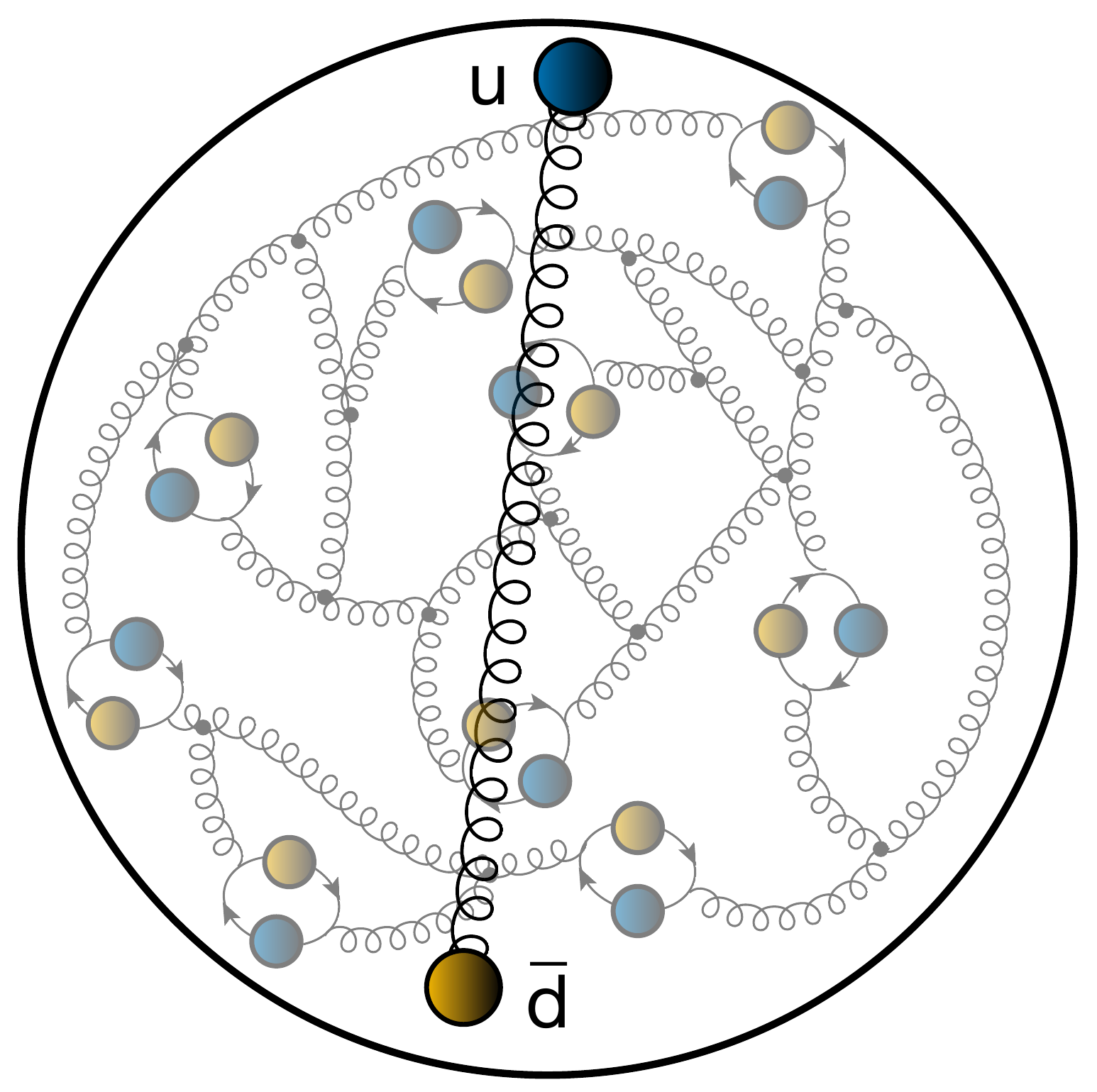}}
\caption{\label{ImagePion}
In contrast to the proton in Fig.\,\ref{Fproton}, the $\pi^+$ contains one valence $u$-quark, one valence $\bar d$-quark; yet, owing to the strong-interaction, it also holds infinitely many gluons and sea quarks.
($\pi^-$ is $d\bar u$ and $\pi^0$ is $u \bar u - d\bar d$.)
}
\end{figure}

\section{Pion and Kaon Charge Radii}
Comparing Fig.\,\ref{ImagePion} with Fig.\,\ref{Fproton}, it is plain that mesons and baryons are distinct forms of hadron matter.  Indeed, the pion ($\pi$) is Nature's lightest hadron \cite{Zyla:2020zbs}; and with a mass similar to that of the $\mu$-lepton, the $\pi$ can seem unnaturally light.
Modern theory explains the pion as simultaneously both a pseudoscalar light-quark+antiquark bound-state and, in the absence of Higgs-boson couplings into QCD, the massless Nambu-Goldstone boson which emerges as a consequence of dynamical chiral symmetry breaking.  This dichotomy is readily explained so long as one employs a symmetry-preserving treatment of QCD's quantum field equations \cite{Lane:1974he, Politzer:1976tv, Delbourgo:1979me, Maris:1997hd, Brodsky:2010xf, Brodsky:2012ku, Qin:2020jig}.  Such proofs reveal that pion observables provide a very clean window onto EHM \cite{Horn:2016rip, Aguilar:2019teb, Chen:2020ijn, Anderle:2021wcy, Roberts:2021nhw, Arrington:2021biu}.

Although the $\pi$ was predicted and discovered more than seventy years ago \cite{Yukawa:1935xg, Lattes:1947mw}, very little has empirically been revealed about its structure \cite{Chang:2021utv, Chang:2022jri, Lu:2022cjx}.  This is primarily because pions are unstable.  For instance, $\pi^\pm$ decay via weak interactions characterised by a leptonic decay constant $f_\pi = 0.0921(8)\,$GeV with a lifetime of just $26.033(5)\,$ns \cite{Zyla:2020zbs}.
Herein, our focus is the electric radius of the charged pion.

With results in hand for the charged-pion elastic form factor, $F_\pi(Q^2)$, then Eq.\,\eqref{EqRadius} yields the pion radius, $r_\pi$.  Many model and QCD theory calculations of $r_\pi$, $f_\pi$ are available.  They show, \emph{inter alia}, that on a large domain of meson mass, the product of radius, $r_{0^-}$, and decay constant, $f_{0^-}$, for ground-state pseudoscalar mesons is practically constant \cite[Fig.\,2A]{Chen:2018rwz}:
\begin{equation}
\label{Eqfpirpi}
f_{0^-} r_{0^-} \approx {\rm constant} \,|\; 0 \leq m_{0^-} \lesssim 1\,{\rm GeV}.
\end{equation}
This feature is a manifestation of EHM, expressed in pseudoscalar meson Bethe-Salpeter wave functions.

The pion form factor has been measured in pion+electron $(\pi e)$ elastic scattering at low-$Q^2$ \cite{Amendolia:1984nz, Amendolia:1986wj, Dally:1982zk, GoughEschrich:2001ji}.  Extractions of $r_\pi$ therefrom are broadly compatible; but the uncertainty is large, \emph{e.g}., $m_\pi$ is known with more than $10\,000$-times better precision.  Given the pion's fundamental role in binding nuclei and its direct links with EHM, this is unsatisfactory; especially since $m_\pi$ and $r_\pi$ are also correlated \cite[Fig.\,2]{Aguilar:2019teb}, entailing that precise values of both are necessary to set meaningful benchmarks for theory.

Today, one typically sees \cite{Zyla:2020zbs}:
\begin{equation}
\label{PDFradius}
r_\pi/{\rm fm} = 0.659 \pm 0.004\,,
\end{equation}
a value obtained from the analysis of data in Refs.\,\cite{Amendolia:1986wj, Dally:1982zk, GoughEschrich:2001ji},
%
%
supplemented by information from $e^+  e^- \to \pi^+ \pi^-$ reactions \cite{Ananthanarayan:2017efc, Colangelo:2018mtw}.  This value has dropped $1.5\,\sigma$ from that quoted two years earlier \cite{Tanabashi:2018oca}: $r_\pi = 0.672(8)\,$fm.  So, it is worth reconsidering available $F_\pi(Q^2)$ data, using the SPM to obtain a fresh $r_\pi$ determination.

\begin{figure}[t]
\centerline{%
\includegraphics[clip, width=0.45\textwidth]{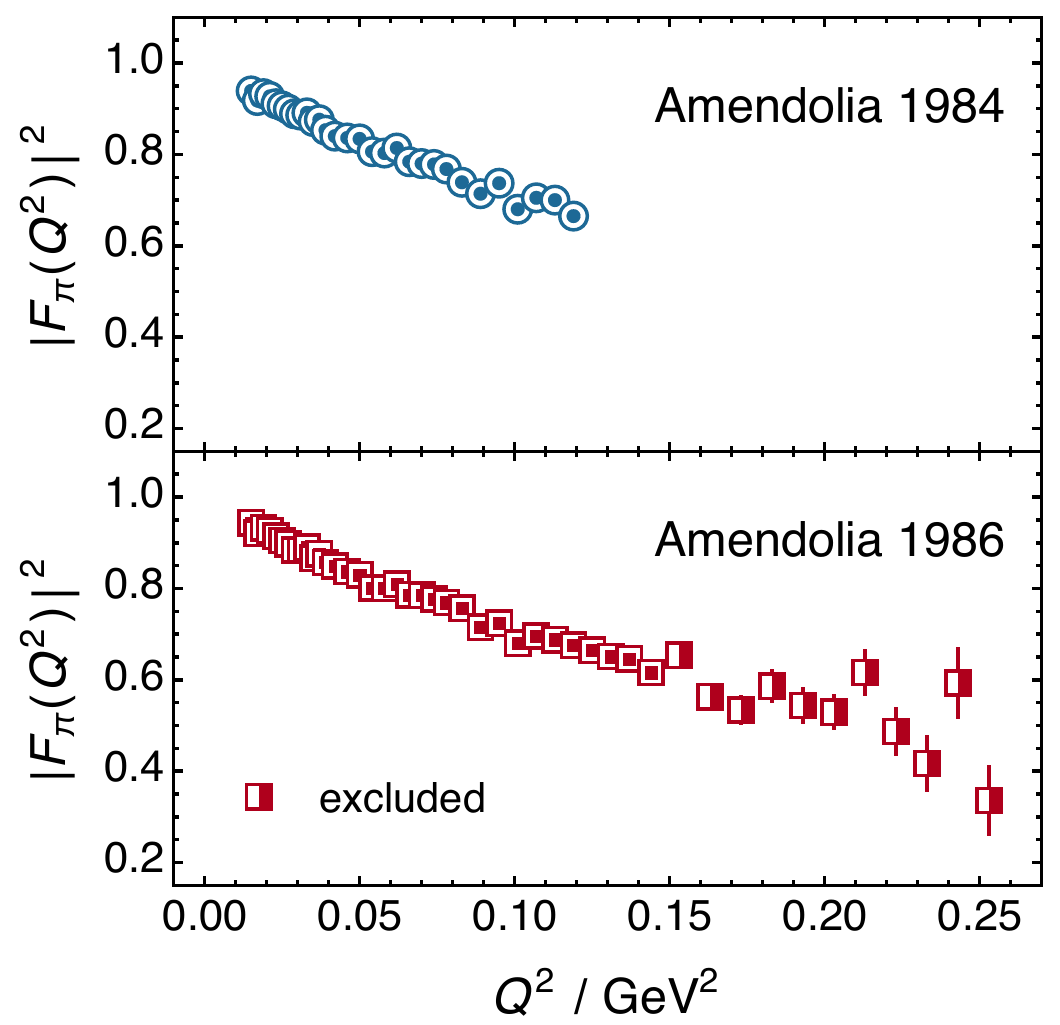}}
\caption{\label{FAmendolia}
NA7 pion form factor data: \emph{upper panel} -- \cite{Amendolia:1984nz}; \emph{lower panel} -- \cite{Amendolia:1986wj}.  Although collected thirty-five years ago, these remain the most dense and extensive sets of low-$Q^2$ data available.
Concerning the 1986 set \cite{Amendolia:1986wj} -- \emph{lower panel}, the SPM analysis in \cite{Cui:2021aee} uses only the square-within-square data.  The half-shaded data are too noisy to add information.
}
\end{figure}

The densest, most extensive sets of low-$Q^2$ data were collected by the NA7 Collaboration \cite{Amendolia:1984nz, Amendolia:1986wj}.  They are depicted in Fig.\,\ref{FAmendolia}.
Within mutual uncertainties, the $r_\pi$ value published in Ref.\,\cite{GoughEschrich:2001ji} agrees with those reported in Ref.\,\cite{Amendolia:1984nz, Amendolia:1986wj}.  However, referred to the results quoted in Ref.\,\cite{Amendolia:1986wj}, the central value in Ref.\,\cite{GoughEschrich:2001ji} is over $2\sigma$ smaller.

Reviewing available low-$Q^2$ $\pi e \to  \pi e$ data \cite{Cui:2021aee}, only the NA7 sets were judged worth considering for reanalysis using the SPM.
The $Q^2$-coverage, density, and precision of the data in Ref.\,\cite{Dally:1982zk} are insufficient to contribute anything beyond what is already available in the NA7 sets;
and Ref.\,\cite{GoughEschrich:2001ji} reports only a radius obtained by assuming a monopole form factor in a fit to unpublished data.

As usual, the SPM analysis follows the pattern established by the $r_p$ study, Secs.\,\ref{SecTheory}, \ref{Recon}, using \cite{Cui:2021aee} the entire body of $N=30$ 1984 NA7 data on $0.015\leq Q^2/{\rm GeV}^2\leq 0.119$ \cite{Amendolia:1984nz} and the $N=35$ lowest-$Q^2$ points from the 1986 NA7 set \cite{Amendolia:1986wj}, $0.015\leq Q^2/{\rm GeV}^2\leq 0.144$.
Here, $6 \leq M \leq N/2$, delivering ${\cal O}(10^5 - 10^7)$ or ${\cal O}(10^6 - 10^9)$ possible interpolators for the 1984 and 1986 data, respectively.
Steps (i)--(iv) in Table~\ref{TabSPMprocedure} are implemented within these spaces \cite{Cui:2021aee}; and as with $r_p$, $r_p^{\rm mag}$, the optimal smoothing parameter $\epsilon \simeq 0$ in all $r_\pi$-related cases.

\begin{figure}[t]
\centerline{%
\includegraphics[clip, width=0.42\textwidth]{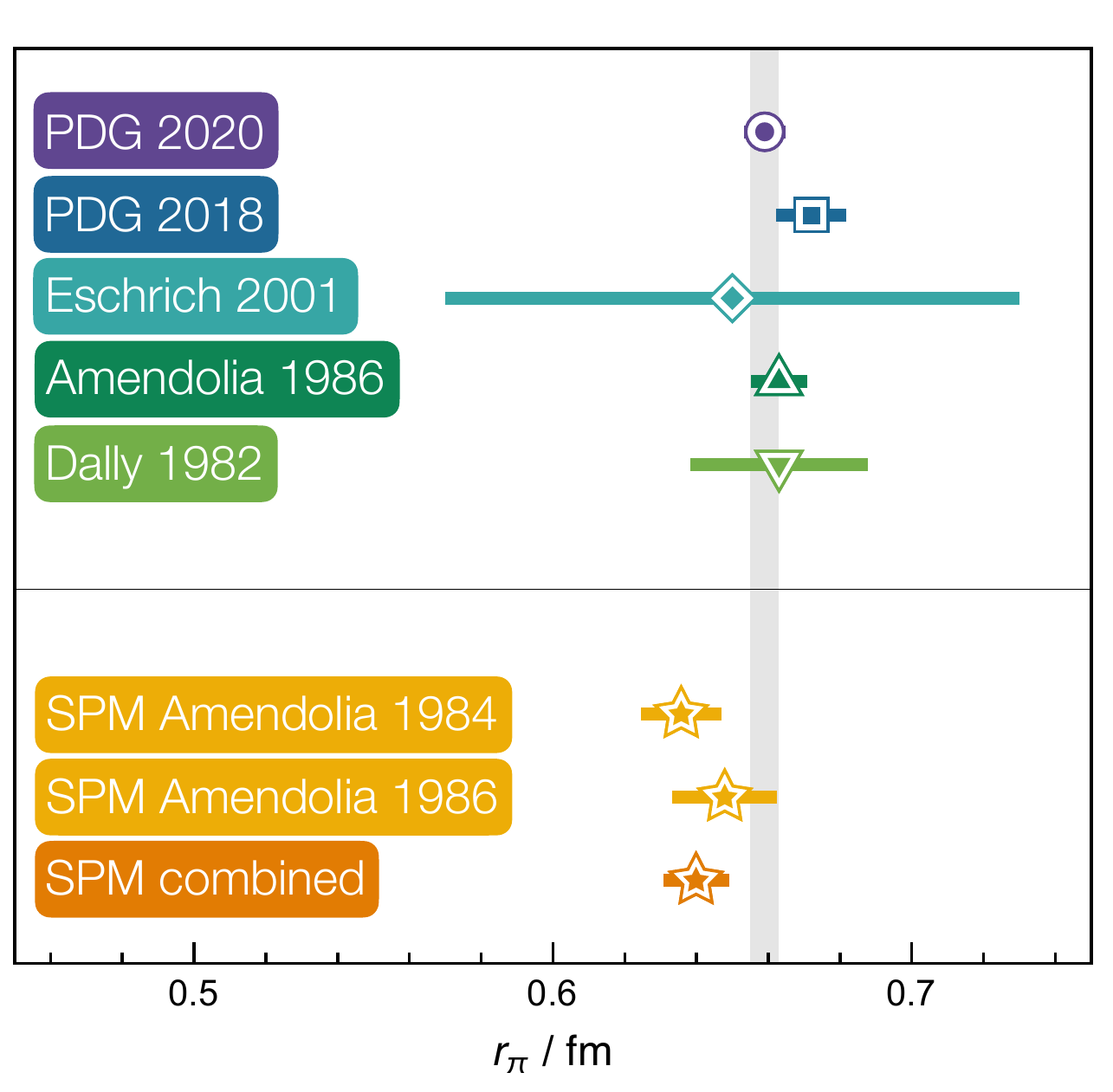}}
\caption{\label{Fradpi}
Pion charge radius:
purple circle (PDG, 2020) \cite{Zyla:2020zbs};
blue square (PDG 2018) \cite{Tanabashi:2018oca};
cyan diamond \cite{GoughEschrich:2001ji};
green up-triangle \cite{Amendolia:1986wj};
light-green down-triangle \cite{Dally:1982zk}.
The vertical grey band marks the uncertainty in Eq.\,\eqref{PDFradius}.
SPM results  \cite{Cui:2021aee} -- gold stars:
SPM A-84 from \cite{Amendolia:1984nz};
SPM A-86 from \cite{Amendolia:1986wj}; and orange star -- SPM combined, the average in Eq.\,\eqref{rpiNA7}.
}
\end{figure}

Subjecting both sets of NA7 data to an array of robustness tests, like those detailed in Sec.\,\ref{SecDeuteron} below, it was found \cite{Cui:2021aee} that with $M \in {\cal O}_M=\{ M_{ji} = 2 + 4 j + i\,\vert\, j=1,2,3,i=1,2,3\}$, one obtains neither a Gaussian distribution nor recovers the input radius.  Given experience with $r_p$ and $r_p^{\rm mag}$, the appearance of such exceptional cases can be linked to the character of the NA7 data.  On the other hand, all robustness tests were satisfied for $M\in {\cal S}_M=\{M_j=2+4j\,\vert\ j = 1,2,3,4\}$.  Hence, these values were used to extract $r_\pi$:
\begin{subequations}
\label{pionradius}
\begin{align}
\label{rpi84}
\,_{\rm SPM}r_\pi^{\rm 84}/{\rm fm} & = 0.636 \pm 0.009_{\rm stat}\,,\\
\label{rpi86}
\,_{\rm SPM}r_\pi^{\rm 86}/{\rm fm} & = 0.648 \pm 0.013_{\rm stat}\,,
\end{align}
\end{subequations}
which are mutually consistent.  Notably, the SPM result is $\approx 1\,\sigma$ below the error-weighted average of each of the original function-choice dependent determinations: $\bar r_\pi^{84}=0.650(8)\,$fm and $\bar r_\pi^{86}=0.659(5)\,$fm.
The uncertainty-weighted average of the radii in Eqs.\,\eqref{pionradius} is
\begin{equation}
\label{rpiNA7}
\,_{\rm SPM}r_\pi^{\rm NA7} = 0.640 \pm 0.007_{\rm stat}\;{\rm fm}\,.
\end{equation}
The results in Eqs.\,\eqref{pionradius}, \eqref{rpiNA7} are compared with other extractions in Fig.\,\ref{Fradpi}: the SPM reanalysis supports the recent downward shift of the PDG average.

Since extant $\pi e$ elastic scattering data do not match analogous modern $ep$ data in precision, density of coverage, or low-$Q^2$ reach, it is today sensible to refrain from judging the true size of $r_\pi$.  A final determination must await improved $\pi e$ data.

\begin{figure}[t]
\centerline{%
\includegraphics[clip, width=0.49\textwidth]{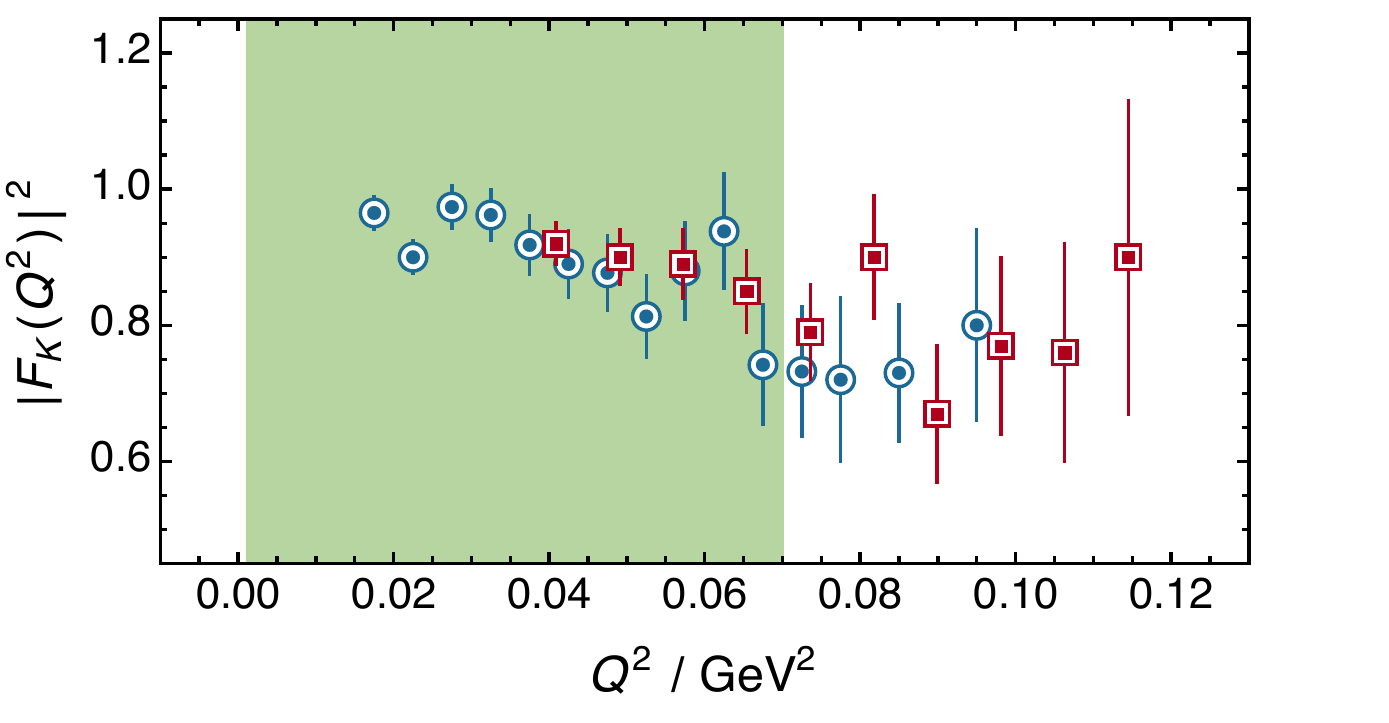}}
\caption{\label{FradK}
Existing low-$Q^2$ kaon form factor data (red squares \cite{Dally:1980dj} and blue circles \cite{Amendolia:1986ui}).  Simple monopole-squared fits to this data are used for existing inferences of the charged-kaon electric radius \cite{Zyla:2020zbs}.
The shaded region indicates the anticipated expansion of $Q^2$ coverage achievable with high-luminosity $K$ beams at a new QCD facility \cite{Adams:2018pwt, Andrieux:2020}.
}
\end{figure}

Suppressing Higgs couplings into QCD, the kaon and pion would be identical; hence,  differences between the $K$ and $\pi$ express Higgs-boson modulations of EHM, \emph{viz}.\ interference between Nature's two known sources of mass.  One of the simplest examples is the ratio of kaon and pion charge radii.  However, the charged-kaon radius, $r_K$, is very poorly constrained.  The commonly quoted value is obtained \cite{Zyla:2020zbs} by averaging the results of separate monopole-squared fits to the distinct data sets \cite{Dally:1980dj, Amendolia:1986ui} depicted in Fig.\,\ref{FradK}.
Employing the SPM to analyse these same data, one finds \cite{Cui:2021aee} that the largest possible roughness penalty is returned, $\epsilon = 1$.  This means that the data contain insufficient information to yield an objective result for the radius: the value extracted is strongly influenced by the practitioner's choice of fitting form.
Hence, $r_K$ should be considered known with a precision that is not better than 10\% \cite{Cui:2021aee}.

A new QCD facility at CERN promises to deliver high-luminosity $\pi$ and $K$ beams \cite{Adams:2018pwt, Andrieux:2020}.  Thus, one may expect the next decade to deliver the first new $\pi e$ and $K e$ elastic scattering data in over thirty years, with the anticipated $Q^2$ coverage indicated by the shaded area in Fig.\,\ref{FradK}.  Should such data on this expanded domain meet the stringent demands of density and precision that have above been established as necessary, then they will enable the first truly objective determinations of pion and kaon charge radii.


\section{Deuteron Charge Radius}
\label{SecDeuteron}
\begin{figure}[t]
\centerline{%
\includegraphics[clip, width=0.40\textwidth]{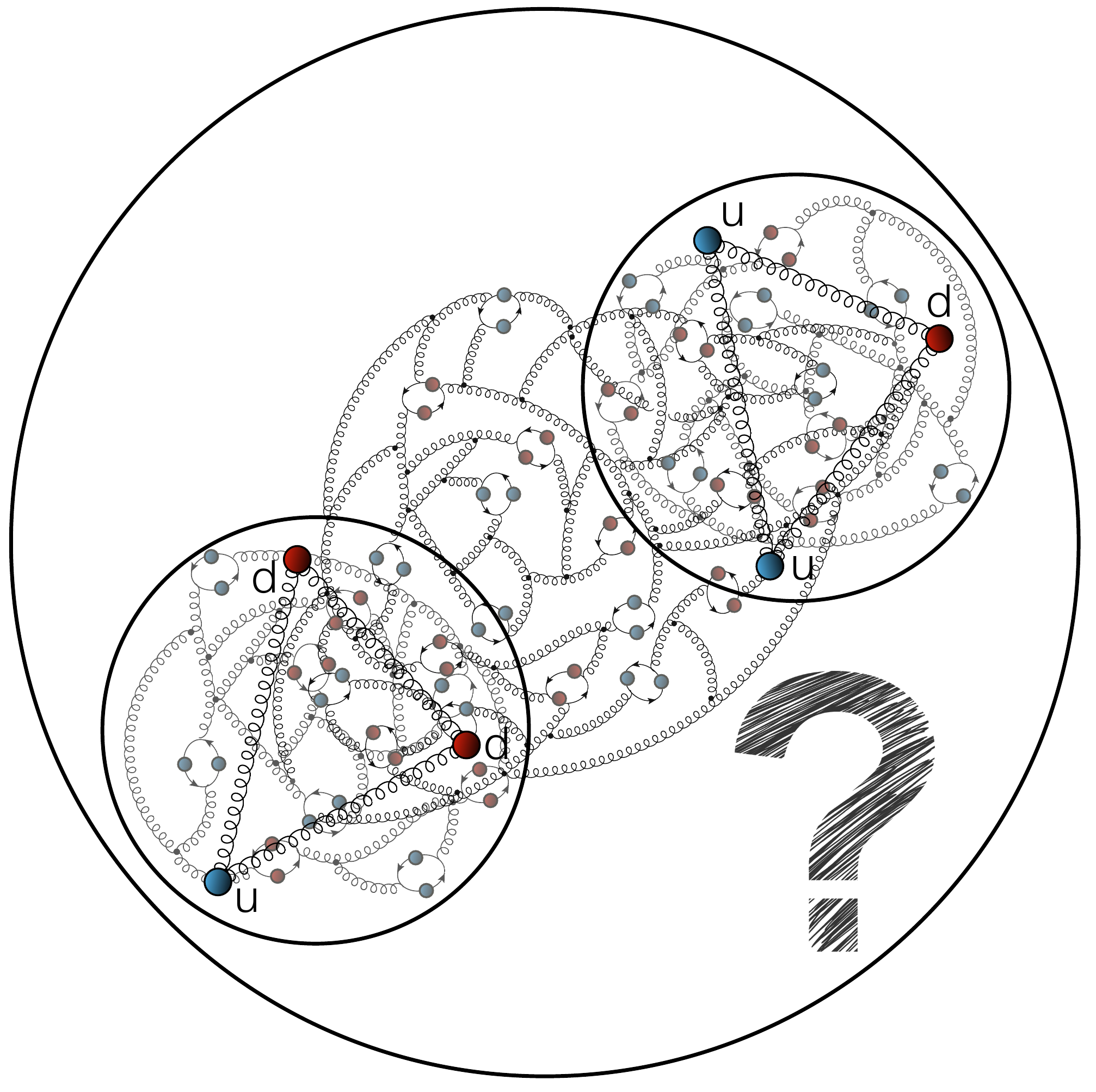}}
\caption{\label{Fdeuteron}
In terms of QCD's Lagrangian quanta, this image might be a fair sketch of the deuteron; but since no realistic QCD-connected calculations of deuteron structure are available, it is currently impossible to know with any confidence.
}
\end{figure}

The deuteron ($d$) is the simplest compound nucleus in Nature: as sketched in Fig.\,\ref{Fdeuteron}, a neutron and proton, somehow bound by SM strong interactions.  Detailed nuclear theory studies of the deuteron exist.  As illustrated elsewhere \cite{Adams:2020aax}, they typically treat the neutron and proton as pointlike and bind them by some model nucleon+nucleon potential.  No realistic QCD-connected calculations of deuteron structure are available.

Notwithstanding the absence of SM theory, the deuteron mass is known with great precision \cite{Tiesinga:2021myr}: the relative uncertainty is $3.0 \times 10^{-10}$.
On the other hand, repeating the stories told above, $r_d$, the deuteron radius, is far less well known.  Extracted from $ed$ elastic scattering, the value usually quoted is $r_d = 2.111(19)\,$fm \cite{Tiesinga:2021myr}.
However, the value $r_d = 2.12562(78)\,$fm is reported from measurements of $2P \to 2S$ transitions in $\mu D$ atoms \cite{CREMA:2016idx}.  This is 25-times more precise than the $ed$ scattering value and, referred to its own error, $19\sigma$ larger.
Further clouding the issue, an analogous measurement using standard atomic deuterium returns $r_d = 2.1415(45)\,$fm \cite{Pohl:2016glp}, which is $3.5\sigma$ above the $\mu D$ value.

\begin{figure*}[!t]
\centerline{\includegraphics[width=0.95\linewidth]{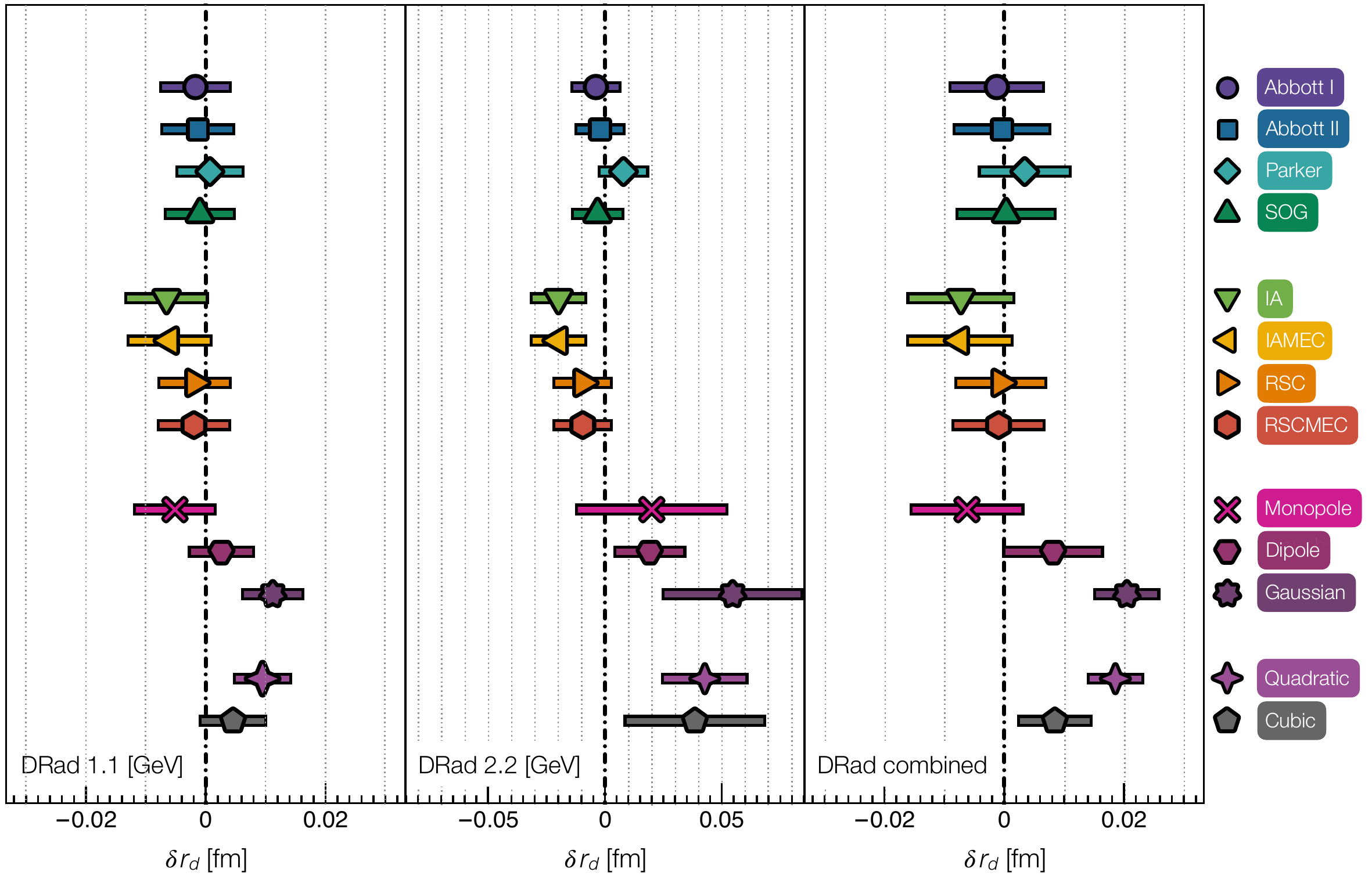}}
	\caption{\label{FValidate}
Bias, $\delta r_{{d}}$, and associated standard error, $\sigma_r$, for the SPM extraction of $r_d$ from $1\,000$ data replicas generated using the thirteen models discussed in Sec.\,\ref{SecDeuteron} on the kinematic domain and with uncertainties anticipated in \cite[DRad]{Zhou:2020cdt}.
%
}
\end{figure*}

Within the context established by this level of disagreement, a new experiment has been proposed at Jefferson Lab \cite[DRad]{JlabDRad}.  Based upon the PRad experimental configuration, DRad aims for an extraction of $r_d$ from $e d$ elastic scattering with precision better than 0.25\%.
Supporting that effort and following the PRad approach, Ref.\,\cite{Zhou:2020cdt} reports a study that uses four parametrisations of the deuteron charge form factor, $G_C^d$, with the aim of identifying fitting functions that can robustly extract $r_d$ from precise data.
As highlighted by the examples drawn above and hereafter shown explicitly, the statistical SPM provides a powerful alternative.

Our illustration proceeds by following the SPM validation approach employed in Refs.\,\cite{Cui:2021vgm, Cui:2021aee, Cui:2021skn}.  Namely, suppose experimental data on $G_C^d$ can be obtained with similar precision and $Q^2$-coverage as PRad proton data.  Then, choose thirteen different models for $G_C^d$, each of which has its own associated value of $r_d$.
The selected models fall into two classes: (\emph{I}) four parametrisations of available $e d$ scattering data; and (\emph{II}) nine theory-based models, ranging from elementary to elaborate.  All correspond to a deuteron with unit electric charge.
Class (\emph{I}) is the collection of parametrising functions used in Ref.\,\cite{Zhou:2020cdt}: \cite[Abbott1]{JLABt20:2000qyq}; \cite[Abbott2]{Kobushkin:1994ed}; \cite[Parker]{asia_parker_2020_4074281}; and \cite[SOG]{Sick:1974suq, Zhou2020}.
The simple models in Class (\emph{II}) are a monopole, a dipole, and a Gaussian, each defined by $r_d=2.1\,$fm; and a quadratic and a cubic polynomial, with their own distinct radii, determined in fits to a model treatment of neutron+proton scattering \cite{Gross:2019thk}.
The elaborate models in Class (\emph{II}) are drawn from the analyses in Ref.\,\cite{Hummel:1990zz, Hummel:1993fq} and correspond to the following treatments of $e d$ scattering:
relativistic impulse approximation based on a one-boson-exchange nucleon+nucleon potential (IA);
IA augmented by meson exchange currents (IAMEC);
relativistic impulse approximation using a Reid soft core potential (RSC);
and RSC augmented by meson exchange currents (RSCMEC).
The associated form factors are determined via spline interpolations of the corresponding numerical results, which are available elsewhere \cite{Higinbotham2020}
%
%
and are linked with the following radii (in fm): IA, $r_{{d}}=2.15798$; IAMEC, $r_{{d}}=2.15727$; RSC, $r_{{d}}=2.0997$; and RSCMEC, $r_{{d}}=2.09912$.
It is worth noting that all models in Class~(\emph{II}) lead to large $\chi^2$ values if used to fit existing experimental data.

As described in Ref.\,\cite{Zhou:2020cdt}, DRad will emulate PRad kinematics, collecting data on $2 \times 10^{-4}\leq Q^{2}/{\rm GeV}^2 \leq 0.05$ using electron beam energies $1.1$, $2.2\,$GeV, with anticipated bin-by-bin statistical uncertainties in the range $0.02-0.07$\% and systematic uncertainties of $0.06-0.16$\%.
Working with this information, the next phase of the illustration consists in following the steps listed in Table~\ref{TabSPMprocedure}.

\begin{figure*}[!t]
\centerline{\includegraphics[width=0.95\linewidth]{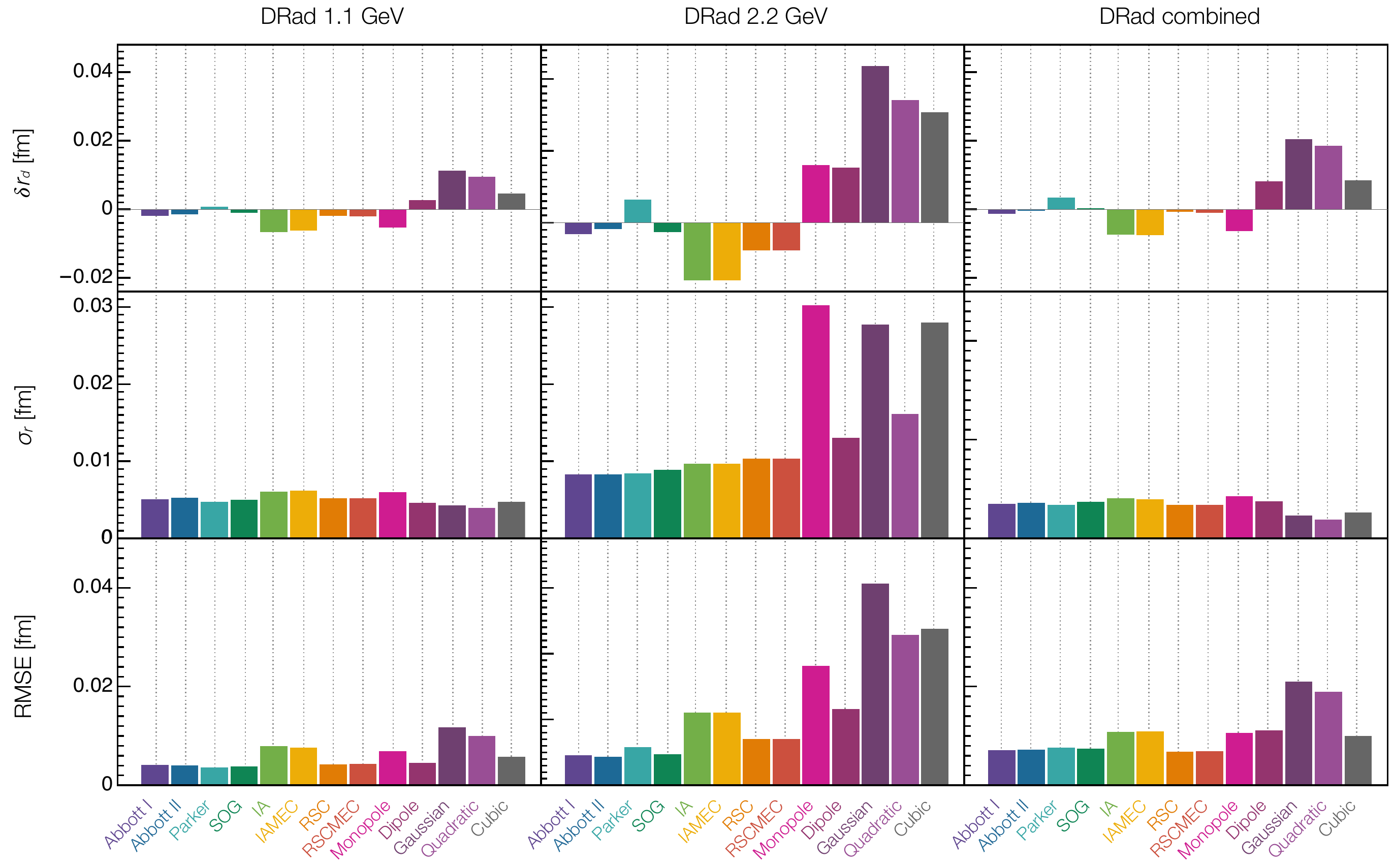}}
	\caption{\label{FValidateX}
Bias $\delta r_{{d}}$, standard error $\sigma_r$ and RMSE of the SPM extrapolation of the deuteron radius for $1\,000$ replicas generated from the thirteen models described in Sec.\,\ref{SecDeuteron}. Notice the limited variation of the RMSE with the generator chosen for the 1.1\,GeV and combined 1.1 and 2.2\,GeV kinematics.
%
}
\end{figure*}


Analysing the results from this procedure enables one to make the following observations.\\[1ex]
\noindent [\textbf{A}] Defining
${\cal S}_M^{1.1}= \{M_j=5+4 j\,|\,j=1,2,3\}$,
${\cal S}_M^{2.2}= \{M_j=2+4 j+i\,|\,j=1,2,3,\, i=1,2\}$,
${\cal S}_M^{1.1 \cup 2.2}= \{M_j=1+4 j+2 i\,|\,j=1,2,3,\, i=1,2\}$,
and considering the different planned DRad electron beam energy settings, $E_\gamma$, then for a given value of $M^{E_\gamma}\in {\cal S}_M^{E_\gamma}$ and all thirteen fitting models, the distribution of SPM-extracted radii is a very clear Gaussian, which is centred on the radius input value and whose characteristics are practically independent of $M$: in these cases, $\sigma_{{\delta\!M}} < \sigma^{{M_j}}_r$.
\\[-1ex]

\noindent [\textbf{B}] Defining the bias: \mbox{$\delta r_M=r_M-r^{\rm input}_M$}, then an assessment of the robustness of the SPM $r_d$ extraction is enabled by Fig.\,\ref{FValidate}.
With beam-energy $1.1\,$GeV (left panel), the SPM delivers a robust radius result, marked by $|\delta r_{{d}}|<\sigma_r$, for all data generators except the Gaussian and quadratic models.
Similarly, for the combined $1.1$ and $2.2\,$GeV data (right panel), the radius determined via the SPM is robust for all data generators except the Gaussian, quadratic and cubic models.
On the other hand, with $2.2\,$GeV beam energy kinematics (central panel, different scale for $\delta r_d$), one finds large biases and errors associated with all five elementary generators; and although the SPM leads to a robust radius determination for the generators based on experimental data, the exposed contrast suggests that analysis of the $2.2\,$GeV kinematic setup alone should be avoided.
\\[-1ex]

\noindent [\textbf{C}] Fig.\,\ref{FValidateX} expands on [\textbf{B}] by quantifying the reliability of SPM analyses of anticipated DRad data using three measures: bias, $\delta r_{d}$; standard deviation, $\sigma_r$; and root mean square error (RMSE)
\begin{equation}
	\mathrm{RMSE}=\sqrt{(\delta r_{\!\s{D}})^2+\sigma_r^2}.
\label{EqRMSE}
\end{equation}
Evidently, for the $1.1\,$GeV (left panel) and combined $1.1$ and $2.2\,$GeV (right panel) kinematics, the RMSE values are practically independent of the generator used to produce the data replicas.  Consequently, the SPM analyses satisfy a standard ``goodness of fit'' criterion \cite{Yan:2018bez}; hence, may objectively be judged to return a reliable expression of the information actually contained in the data.  Considering the middle panel, \emph{i.e}., $2.2$\,GeV beam kinematics alone, the elementary generators are again seen to present a challenge.

It is now worth comparing these remarks and Figs.\,\ref{FValidate}, \ref{FValidateX} with Ref.\,\cite[DRad -- Figs.\,2, 5, 9]{Zhou:2020cdt} and the associated discussion.
The primary conclusion of the latter is that the methods used by the PRad Collaboration to identify ``best fitter'' functions for use in extracting $r_p$ are unsuitable for the $r_d$ case.  So, a new method, tuned specifically to $e d$ scattering, needed to be developed.  It yielded two new best fitter functions, neither of which coincides with any of the forms used in connection with the proton radius extraction.

This contrasts markedly with the SPM validation tests.  As shown above, for practical purposes, all procedures used in connection with the reliable extraction of $r_p$ are equally effective for a SPM extraction of $r_d$: one obtains a robust extraction of the deuteron radius in 85\% (69\%) of cases using DRad $1.1\,$GeV beam ($1.1$+$2.2$ combined) kinematics.  Moreover, the SPM analysis supplies a mathematical demonstration that, used alone, the $2.2\,$GeV beam data cannot deliver an objective result for $r_d$.
Finally, considering the middle row of Fig.\,\ref{FValidateX}, the SPM delivers a mean variance $\bar\sigma_r = 0.0055, 0.0075$ for the $1.1\,$GeV and combined $1.1$+$2.2\,$GeV beam data, respectively, which compares favourably with the mean variance ($0.0035$) obtained using the $ed$-tuned best fitters in Ref.\,\cite[Fig.\,6]{Zhou:2020cdt}.
In applying the SPM to anticipated DRad data and subject to the considerations discussed in this section, one would be using
75-million (45-million with $E_\gamma=2.2\,$GeV excluded) independent deuteron radius values in the determination of $r_d$ via Eq.\,\eqref{SPMrp}.

\section{Outlook}
As a solution is sought for the strong interaction problem within the Standard Model (SM), precise knowledge of the Poincar\'e-invariant radii of hadrons (\emph{e.g}., proton, pion, kaon), and light nuclei (\emph{e.g}., deuteron) has become a high priority.  Yet, recent measurements and new analyses of older data have revealed uncertainties and imprecisions that severely limit the use of these radii as benchmarks for SM theory.  The last decade has revealed that reliable extraction of such radii from electron+hadron scattering requires:
(\emph{i}) precise data, stretching as close as possible to $Q^2=0$, and densely covering a domain $Q^2\lesssim 0.1\,$GeV$^2$; and
(\emph{ii}) elimination of all subjectivity/bias associated with practitioner-dependent choices made in fitting the data and subsequently extrapolating to estimate a $Q^2=0$ derivative.
Herein, we have not addressed requirement (\emph{i}), which is a challenge for facility hardware, and experiment design and conduct.
Regarding requirement (\emph{ii}), different approaches have been suggested, each typically tuned to a given reaction.  We have reviewed an alternative that is applicable in the same form to all cases; namely, the statistical Schlessinger point method (SPM).

The SPM is a tool for interpolating data (broadly defined); and, significantly, from the interpolations, delivering an estimate, with known uncertainty, for the curve underlying that data.  This information may then be used to predict the values of quantities properly determined by data outside the domain of existing measurements.
The SPM's basic strengths are a rigorous mathematical foundation in analytic function theory; the absence of assumptions about theories which may underly the data; the elimination of practitioner-induced bias; and mechanical, uniform implementation in diverse, unrelated contexts, \emph{viz}.\ the absence of a need for system-specific tuning.
With these foundations, the SPM returns an objective expression of the information contained in the data under consideration; and because it makes no reference to theories of physical phenomena, that expression serves as a test for both the experiment and potentially relevant theory.

Whilst it will now be clear, it is nevertheless worth emphasising that the SPM can be used for much more than extracting radii.  The range of existing applications includes
the extrapolation of theory predictions for hadron form factors to large values of $Q^2$, \emph{e.g}., Refs.\,\cite{Chen:2018nsg, Xu:2019ilh, Cui:2020rmu};
continuation from Euclidean to Minkowski space in order to obtain light-front specific quantities from Poincar\'e-covariant wave functions, \emph{e.g}., Refs.\,\cite{Ding:2019qlr, Eichmann:2021vnj};
overcoming the challenge of large mass imbalances between valence degrees-of-freedom in delivering predictions for the semileptonic decays of heavy+light mesons, \emph{e.g}., Ref.\,\cite{Yao:2021pdy};
and the analysis of light-nuclei deep inelastic scattering data \cite{Abrams:2021xum} and its interpretation in terms of parton distribution functions \cite{Cui:2021gzg}.
There is no limit to the scope of the SPM.  Given precise data, densely covering an appropriate domain, it will return an unbiased estimate of a desired observable with an uncertainty that quantitatively reflects the quality of the data and the length of the extrapolation.


\section*{Acknowledgments}
In preparing this article we benefited from constructive interactions with
L.~Chang, O.~Denisov, J.~Friedrich, H.~Gao, D.\,W.~Higinbotham, R.\,J.~Holt, V.~Mokeev, W.-D.~Nowak, C.~Quintans, P.\,E.~Reimer, W.~Xiong and J.~Zhou.
Use of the computer clusters at the Institute for Nonperturbative Physics at Nanjing University is gratefully acknowledged.
Work supported by:
National Natural Science Foundation of China (grant no.\,12135007);
Natural Science Foundation of Jiangsu Province (grant no.\ BK20220323);
and STRONG-2020 ``The strong interaction at the frontier of knowledge: fundamental research and applications'' which received funding from the European Union's Horizon 2020 research and innovation programme (grant no.\,824093).



\begin{thebibliography}{137}
\providecommand{\natexlab}[1]{#1}
\providecommand{\url}[1]{\texttt{#1}}
\providecommand{\urlprefix}{URL }
\expandafter\ifx\csname urlstyle\endcsname\relax
  \providecommand{\doi}[1]{doi:\discretionary{}{}{}#1}\else
  \providecommand{\doi}[1]{doi:\discretionary{}{}{}\begingroup
  \urlstyle{rm}\url{#1}\endgroup}\fi
\providecommand{\bibinfo}[2]{#2}

\bibitem[{Marciano and Pagels(1978)}]{Marciano:1977su}
\bibinfo{author}{W.~J. Marciano}, \bibinfo{author}{H.~Pagels},
  \bibinfo{title}{{Quantum Chromodynamics: A Review}}, \bibinfo{journal}{Phys.
  Rept.} \bibinfo{volume}{36} (\bibinfo{year}{1978}) \bibinfo{pages}{137}.

\bibitem[{Brock et~al.(1995)}]{Brock:1993sz}
\bibinfo{author}{R.~Brock}, et~al., \bibinfo{title}{{Handbook of perturbative
  QCD: Version 1.0}}, \bibinfo{journal}{Rev. Mod. Phys.} \bibinfo{volume}{67}
  (\bibinfo{year}{1995}) \bibinfo{pages}{157--248}.

\bibitem[{Brodsky et~al.(2010)Brodsky, Roberts, Shrock, and
  Tandy}]{Brodsky:2010xf}
\bibinfo{author}{S.~J. Brodsky}, \bibinfo{author}{C.~D. Roberts},
  \bibinfo{author}{R.~Shrock}, \bibinfo{author}{P.~C. Tandy},
  \bibinfo{title}{{New perspectives on the quark condensate}},
  \bibinfo{journal}{Phys. Rev. C} \bibinfo{volume}{82} (\bibinfo{year}{2010})
  \bibinfo{pages}{022201(R)}.

\bibitem[{Zyla et~al.(2020)}]{Zyla:2020zbs}
\bibinfo{author}{P.~Zyla}, et~al., \bibinfo{title}{{Review of Particle
  Physics}}, \bibinfo{journal}{PTEP} \bibinfo{volume}{2020}
  (\bibinfo{year}{2020}) \bibinfo{pages}{083C01}.

\bibitem[{Stodolna et~al.(2013)Stodolna, Rouz\'ee, L\'epine, Cohen, Robicheaux,
  Gijsbertsen, Jungmann, Bordas, and Vrakking}]{PhysRevLett.110.213001}
\bibinfo{author}{A.~S. Stodolna}, \bibinfo{author}{A.~Rouz\'ee},
  \bibinfo{author}{F.~L\'epine}, \bibinfo{author}{S.~Cohen},
  \bibinfo{author}{F.~Robicheaux}, \bibinfo{author}{A.~Gijsbertsen},
  \bibinfo{author}{J.~H. Jungmann}, \bibinfo{author}{C.~Bordas},
  \bibinfo{author}{M.~J.~J. Vrakking}, \bibinfo{title}{Hydrogen Atoms under
  Magnification: Direct Observation of the Nodal Structure of Stark States},
  \bibinfo{journal}{Phys. Rev. Lett.} \bibinfo{volume}{110}
  (\bibinfo{year}{2013}) \bibinfo{pages}{213001}.

\bibitem[{Roberts(2017)}]{Roberts:2016vyn}
\bibinfo{author}{C.~D. Roberts}, \bibinfo{title}{{Perspective on the origin of
  hadron masses}}, \bibinfo{journal}{Few Body Syst.} \bibinfo{volume}{58}
  (\bibinfo{year}{2017}) \bibinfo{pages}{5}.

\bibitem[{Adams et~al.(2018)}]{Adams:2018pwt}
\bibinfo{author}{B.~Adams}, et~al., \bibinfo{title}{{Letter of Intent: A New
  QCD facility at the M2 beam line of the CERN SPS (COMPASS++/AMBER) --
  arXiv:1808.00848 [hep-ex]$\!$}} .

\bibitem[{Andrieux and Parsamyan(2020)}]{Andrieux:2020}
\bibinfo{author}{V.~Andrieux}, \bibinfo{author}{B.~Parsamyan},
  \bibinfo{title}{{From COMPASS to AMBER: exploring fundamental properties of
  hadrons}}, \bibinfo{journal}{CERN EP Newsletter} \bibinfo{volume}{2020/12 -
  2021/02}.

\bibitem[{Horn and Roberts(2016)}]{Horn:2016rip}
\bibinfo{author}{T.~Horn}, \bibinfo{author}{C.~D. Roberts},
  \bibinfo{title}{{The pion: an enigma within the Standard Model}},
  \bibinfo{journal}{J. Phys. G.} \bibinfo{volume}{43} (\bibinfo{year}{2016})
  \bibinfo{pages}{073001}.

\bibitem[{Aguilar et~al.(2019)}]{Aguilar:2019teb}
\bibinfo{author}{A.~C. Aguilar}, et~al., \bibinfo{title}{{Pion and Kaon
  Structure at the Electron-Ion Collider}}, \bibinfo{journal}{Eur. Phys. J. A}
  \bibinfo{volume}{55} (\bibinfo{year}{2019}) \bibinfo{pages}{190}.

\bibitem[{Brodsky et~al.(2020)}]{Brodsky:2020vco}
\bibinfo{author}{S.~J. Brodsky}, et~al., \bibinfo{title}{{Strong QCD from
  Hadron Structure Experiments}}, \bibinfo{journal}{Int. J. Mod. Phys. E}
  \bibinfo{volume}{29}~(\bibinfo{number}{08}) (\bibinfo{year}{2020})
  \bibinfo{pages}{2030006}.

\bibitem[{Chen et~al.(2020)Chen, Guo, Roberts, and Wang}]{Chen:2020ijn}
\bibinfo{author}{X.~Chen}, \bibinfo{author}{F.-K. Guo}, \bibinfo{author}{C.~D.
  Roberts}, \bibinfo{author}{R.~Wang}, \bibinfo{title}{{Selected Science
  Opportunities for the EicC}}, \bibinfo{journal}{Few Body Syst.}
  \bibinfo{volume}{61} (\bibinfo{year}{2020}) \bibinfo{pages}{43}.

\bibitem[{Anderle et~al.(2021)}]{Anderle:2021wcy}
\bibinfo{author}{D.~P. Anderle}, et~al., \bibinfo{title}{{Electron-ion collider
  in China}}, \bibinfo{journal}{Front. Phys. (Beijing)}
  \bibinfo{volume}{16}~(\bibinfo{number}{6}) (\bibinfo{year}{2021})
  \bibinfo{pages}{64701}.

\bibitem[{Arrington et~al.(2021)}]{Arrington:2021biu}
\bibinfo{author}{J.~Arrington}, et~al., \bibinfo{title}{{Revealing the
  structure of light pseudoscalar mesons at the electron\textendash{}ion
  collider}}, \bibinfo{journal}{J. Phys. G} \bibinfo{volume}{48}
  (\bibinfo{year}{2021}) \bibinfo{pages}{075106}.

\bibitem[{Abdul~Khalek et~al.(2021)}]{AbdulKhalek:2021gbh}
\bibinfo{author}{R.~Abdul~Khalek}, et~al., \bibinfo{title}{{Science
  Requirements and Detector Concepts for the Electron-Ion Collider: EIC Yellow
  Report -- \emph{arXiv:2103.05419 [physics.ins-det]}}} .

\bibitem[{Du et~al.(2020)Du, Baru, Guo, Hanhart, Mei\ss{}ner, Nefediev, and
  Strakovsky}]{Du:2020bqj}
\bibinfo{author}{M.-L. Du}, \bibinfo{author}{V.~Baru}, \bibinfo{author}{F.-K.
  Guo}, \bibinfo{author}{C.~Hanhart}, \bibinfo{author}{U.-G. Mei\ss{}ner},
  \bibinfo{author}{A.~Nefediev}, \bibinfo{author}{I.~Strakovsky},
  \bibinfo{title}{{Deciphering the mechanism of near-threshold $J/\psi$
  photoproduction}}, \bibinfo{journal}{Eur. Phys. J. C}
  \bibinfo{volume}{80}~(\bibinfo{number}{11}) (\bibinfo{year}{2020})
  \bibinfo{pages}{1053}.

\bibitem[{Xu et~al.(2021)Xu, Chen, Yao, Binosi, Cui, and Roberts}]{Xu:2021mju}
\bibinfo{author}{Y.-Z. Xu}, \bibinfo{author}{S.~Chen}, \bibinfo{author}{Z.-Q.
  Yao}, \bibinfo{author}{D.~Binosi}, \bibinfo{author}{Z.-F. Cui},
  \bibinfo{author}{C.~D. Roberts}, \bibinfo{title}{{Vector-meson production and
  vector meson dominance}}, \bibinfo{journal}{Eur. Phys. J. C}
  \bibinfo{volume}{81} (\bibinfo{year}{2021}) \bibinfo{pages}{895}.

\bibitem[{Sun et~al.(2022)Sun, Tong, and Yuan}]{Sun:2021pyw}
\bibinfo{author}{P.~Sun}, \bibinfo{author}{X.-B. Tong},
  \bibinfo{author}{F.~Yuan}, \bibinfo{title}{{Near threshold heavy quarkonium
  photoproduction at large momentum transfer}}, \bibinfo{journal}{Phys. Rev. D}
  \bibinfo{volume}{105}~(\bibinfo{number}{5}) (\bibinfo{year}{2022})
  \bibinfo{pages}{054032}.

\bibitem[{Roberts et~al.(2021)Roberts, Richards, Horn, and
  Chang}]{Roberts:2021nhw}
\bibinfo{author}{C.~D. Roberts}, \bibinfo{author}{D.~G. Richards},
  \bibinfo{author}{T.~Horn}, \bibinfo{author}{L.~Chang},
  \bibinfo{title}{{Insights into the emergence of mass from studies of pion and
  kaon structure}}, \bibinfo{journal}{Prog. Part. Nucl. Phys.}
  \bibinfo{volume}{120} (\bibinfo{year}{2021}) \bibinfo{pages}{103883}.

\bibitem[{Carlson et~al.(2006)Carlson, Jaffe, and Wiles}]{millennium:2006}
\bibinfo{editor}{J.~Carlson}, \bibinfo{editor}{A.~Jaffe},
  \bibinfo{editor}{A.~Wiles} (Eds.), \bibinfo{title}{{\emph{The Millenium Prize
  Problems}}}, \bibinfo{publisher}{American Mathematical Society, Providence},
  \bibinfo{year}{2006}.

\bibitem[{Abe et~al.(2017)}]{Super-Kamiokande:2016exg}
\bibinfo{author}{K.~Abe}, et~al., \bibinfo{title}{{Search for proton decay via
  $p \to e^+\pi^0$ and $p \to \mu^+\pi^0$ in 0.31
  megaton\textperiodcentered{}years exposure of the Super-Kamiokande water
  Cherenkov detector}}, \bibinfo{journal}{Phys. Rev. D}
  \bibinfo{volume}{95}~(\bibinfo{number}{1}) (\bibinfo{year}{2017})
  \bibinfo{pages}{012004}.

\bibitem[{Hei{\ss}e et~al.(2019)Hei{\ss}e, Rau, K\"ohler-Langes, Quint, Werth,
  Sturm, and Blaum}]{Heisse:2019xnz}
\bibinfo{author}{F.~Hei{\ss}e}, \bibinfo{author}{S.~Rau},
  \bibinfo{author}{F.~K\"ohler-Langes}, \bibinfo{author}{W.~Quint},
  \bibinfo{author}{G.~Werth}, \bibinfo{author}{S.~Sturm},
  \bibinfo{author}{K.~Blaum}, \bibinfo{title}{{High-precision mass spectrometer
  for light ions}}, \bibinfo{journal}{Phys. Rev. A}
  \bibinfo{volume}{100}~(\bibinfo{number}{2}) (\bibinfo{year}{2019})
  \bibinfo{pages}{022518}.

\bibitem[{Brodsky et~al.(2015)Brodsky, de~Teramond, Dosch, and
  Erlich}]{Brodsky:2014yha}
\bibinfo{author}{S.~J. Brodsky}, \bibinfo{author}{G.~F. de~Teramond},
  \bibinfo{author}{H.~G. Dosch}, \bibinfo{author}{J.~Erlich},
  \bibinfo{title}{{Light-Front Holographic QCD and Emerging Confinement}},
  \bibinfo{journal}{Phys. Rept.} \bibinfo{volume}{584} (\bibinfo{year}{2015})
  \bibinfo{pages}{1--105}.

\bibitem[{Giannini and Santopinto(2015)}]{Giannini:2015zia}
\bibinfo{author}{M.~M. Giannini}, \bibinfo{author}{E.~Santopinto},
  \bibinfo{title}{{The hypercentral Constituent Quark Model and its application
  to baryon properties}}, \bibinfo{journal}{Chin. J. Phys.}
  \bibinfo{volume}{53} (\bibinfo{year}{2015}) \bibinfo{pages}{020301}.

\bibitem[{Eichmann et~al.(2016)Eichmann, Sanchis-Alepuz, Williams, Alkofer, and
  Fischer}]{Eichmann:2016yit}
\bibinfo{author}{G.~Eichmann}, \bibinfo{author}{H.~Sanchis-Alepuz},
  \bibinfo{author}{R.~Williams}, \bibinfo{author}{R.~Alkofer},
  \bibinfo{author}{C.~S. Fischer}, \bibinfo{title}{{Baryons as relativistic
  three-quark bound states}}, \bibinfo{journal}{Prog. Part. Nucl. Phys.}
  \bibinfo{volume}{91} (\bibinfo{year}{2016}) \bibinfo{pages}{1--100}.

\bibitem[{Qin et~al.(2019)Qin, Roberts, and Schmidt}]{Qin:2019hgk}
\bibinfo{author}{S.-X. Qin}, \bibinfo{author}{C.~D. Roberts},
  \bibinfo{author}{S.~M. Schmidt}, \bibinfo{title}{{Spectrum of light- and
  heavy-baryons}}, \bibinfo{journal}{Few Body Syst.} \bibinfo{volume}{60}
  (\bibinfo{year}{2019}) \bibinfo{pages}{26}.

\bibitem[{Qin and Roberts(2021)}]{Qin:2020jig}
\bibinfo{author}{S.-X. Qin}, \bibinfo{author}{C.~D. Roberts},
  \bibinfo{title}{{Resolving the Bethe-Salpeter kernel}},
  \bibinfo{journal}{Chin. Phys. Lett. \emph{Express}}
  \bibinfo{volume}{38}~(\bibinfo{number}{7}) (\bibinfo{year}{2021})
  \bibinfo{pages}{071201}.

\bibitem[{Yin et~al.(2021)Yin, Cui, Roberts, and Segovia}]{Yin:2021uom}
\bibinfo{author}{P.-L. Yin}, \bibinfo{author}{Z.-F. Cui},
  \bibinfo{author}{C.~D. Roberts}, \bibinfo{author}{J.~Segovia},
  \bibinfo{title}{{Masses of positive- and negative-parity hadron
  ground-states, including those with heavy quarks}}, \bibinfo{journal}{Eur.
  Phys. J. C} \bibinfo{volume}{81}~(\bibinfo{number}{4}) (\bibinfo{year}{2021})
  \bibinfo{pages}{327}.

\bibitem[{Frisch and Stern(1933)}]{FrischStern1}
\bibinfo{author}{R.~Frisch}, \bibinfo{author}{O.~Stern},
  \bibinfo{title}{{\"Uber die magnetische Ablenkung von Wasserstoffmolek\"ulen
  und das magnetische Moment des Protons. I}}, \bibinfo{journal}{Z. Physik}
  \bibinfo{volume}{85} (\bibinfo{year}{1933}) \bibinfo{pages}{4--16}.

\bibitem[{Hofstadter(1956)}]{Hofstadter:1956qs}
\bibinfo{author}{R.~Hofstadter}, \bibinfo{title}{{Electron scattering and
  nuclear structure}}, \bibinfo{journal}{Rev. Mod. Phys.} \bibinfo{volume}{28}
  (\bibinfo{year}{1956}) \bibinfo{pages}{214--254}.

\bibitem[{Eides et~al.(2001)Eides, Grotch, and Shelyuto}]{Eides:2000xc}
\bibinfo{author}{M.~I. Eides}, \bibinfo{author}{H.~Grotch},
  \bibinfo{author}{V.~A. Shelyuto}, \bibinfo{title}{{Theory of light hydrogen -
  like atoms}}, \bibinfo{journal}{Phys. Rept.} \bibinfo{volume}{342}
  (\bibinfo{year}{2001}) \bibinfo{pages}{63--261}.

\bibitem[{Schlessinger(1968)}]{PhysRev.167.1411}
\bibinfo{author}{L.~Schlessinger}, \bibinfo{title}{Use of Analyticity in the
  Calculation of Nonrelativistic Scattering Amplitudes},
  \bibinfo{journal}{Phys. Rev.} \bibinfo{volume}{167} (\bibinfo{year}{1968})
  \bibinfo{pages}{1411--1423}.

\bibitem[{Schlessinger and Schwartz(1966)}]{Schlessinger:1966zz}
\bibinfo{author}{L.~Schlessinger}, \bibinfo{author}{C.~Schwartz},
  \bibinfo{title}{{Analyticity as a Useful Computation Tool}},
  \bibinfo{journal}{Phys. Rev. Lett.} \bibinfo{volume}{16}
  (\bibinfo{year}{1966}) \bibinfo{pages}{1173--1174}.

\bibitem[{Tripolt et~al.(2017)Tripolt, Haritan, Wambach, and
  Moiseyev}]{Tripolt:2016cya}
\bibinfo{author}{R.~A. Tripolt}, \bibinfo{author}{I.~Haritan},
  \bibinfo{author}{J.~Wambach}, \bibinfo{author}{N.~Moiseyev},
  \bibinfo{title}{{Threshold energies and poles for hadron physical problems by
  a model-independent universal algorithm}}, \bibinfo{journal}{Phys. Lett. B}
  \bibinfo{volume}{774} (\bibinfo{year}{2017}) \bibinfo{pages}{411--416}.

\bibitem[{Chen et~al.(2019)Chen, Lu, Binosi, Roberts, Rodr\'\i{}guez-Quintero,
  and Segovia}]{Chen:2018nsg}
\bibinfo{author}{C.~Chen}, \bibinfo{author}{Y.~Lu},
  \bibinfo{author}{D.~Binosi}, \bibinfo{author}{C.~D. Roberts},
  \bibinfo{author}{J.~Rodr\'\i{}guez-Quintero}, \bibinfo{author}{J.~Segovia},
  \bibinfo{title}{{Nucleon-to-Roper electromagnetic transition form factors at
  large $Q^2$}}, \bibinfo{journal}{Phys. Rev. D} \bibinfo{volume}{99}
  (\bibinfo{year}{2019}) \bibinfo{pages}{034013}.

\bibitem[{Bernauer et~al.(2010)}]{Bernauer:2010wm}
\bibinfo{author}{J.~C. Bernauer}, et~al., \bibinfo{title}{{High-precision
  determination of the electric and magnetic form factors of the proton}},
  \bibinfo{journal}{Phys. Rev. Lett.} \bibinfo{volume}{105}
  (\bibinfo{year}{2010}) \bibinfo{pages}{242001}.

\bibitem[{Pohl et~al.(2010)}]{Pohl:2010zza}
\bibinfo{author}{R.~Pohl}, et~al., \bibinfo{title}{{The size of the proton}},
  \bibinfo{journal}{Nature} \bibinfo{volume}{466} (\bibinfo{year}{2010})
  \bibinfo{pages}{213--216}.

\bibitem[{Zhan et~al.(2011)}]{Zhan:2011ji}
\bibinfo{author}{X.~Zhan}, et~al., \bibinfo{title}{{High-Precision Measurement
  of the Proton Elastic Form Factor Ratio $\mu_pG_E/G_M$ at low $Q^2$}},
  \bibinfo{journal}{Phys. Lett. B} \bibinfo{volume}{705} (\bibinfo{year}{2011})
  \bibinfo{pages}{59--64}.

\bibitem[{Antognini et~al.(2013)}]{Antognini:1900ns}
\bibinfo{author}{A.~Antognini}, et~al., \bibinfo{title}{{Proton Structure from
  the Measurement of $2S-2P$ Transition Frequencies of Muonic Hydrogen}},
  \bibinfo{journal}{Science} \bibinfo{volume}{339} (\bibinfo{year}{2013})
  \bibinfo{pages}{417--420}.

\bibitem[{Beyer et~al.(2017)}]{Beyer:2017gug}
\bibinfo{author}{A.~Beyer}, et~al., \bibinfo{title}{{The Rydberg constant and
  proton size from atomic hydrogen}}, \bibinfo{journal}{Science}
  \bibinfo{volume}{358}~(\bibinfo{number}{6359}) (\bibinfo{year}{2017})
  \bibinfo{pages}{79--85}.

\bibitem[{Fleurbaey et~al.(2018)Fleurbaey, Galtier, Thomas, Bonnaud, Julien,
  Biraben, Nez, Abgrall, and Gu{\'e}na}]{Fleurbaey:2018fih}
\bibinfo{author}{H.~Fleurbaey}, \bibinfo{author}{S.~Galtier},
  \bibinfo{author}{S.~Thomas}, \bibinfo{author}{M.~Bonnaud},
  \bibinfo{author}{L.~Julien}, \bibinfo{author}{F.~Biraben},
  \bibinfo{author}{F.~Nez}, \bibinfo{author}{M.~Abgrall},
  \bibinfo{author}{J.~Gu{\'e}na}, \bibinfo{title}{{New Measurement of the
  $1S-3S$ Transition Frequency of Hydrogen: Contribution to the Proton Charge
  Radius Puzzle}}, \bibinfo{journal}{Phys. Rev. Lett.} \bibinfo{volume}{120}
  (\bibinfo{year}{2018}) \bibinfo{pages}{183001}.

\bibitem[{Mihovilovi\v{c} et~al.(2021)}]{Mihovilovic:2019jiz}
\bibinfo{author}{M.~Mihovilovi\v{c}}, et~al., \bibinfo{title}{{The proton
  charge radius extracted from the initial-state radiation experiment at
  MAMI}}, \bibinfo{journal}{Eur. Phys. J. A}
  \bibinfo{volume}{57}~(\bibinfo{number}{3}) (\bibinfo{year}{2021})
  \bibinfo{pages}{107}.

\bibitem[{Bezginov et~al.(2019)Bezginov, Valdez, Horbatsch, Marsman, Vutha, and
  Hessels}]{Bezginov1007}
\bibinfo{author}{N.~Bezginov}, \bibinfo{author}{T.~Valdez},
  \bibinfo{author}{M.~Horbatsch}, \bibinfo{author}{A.~Marsman},
  \bibinfo{author}{A.~C. Vutha}, \bibinfo{author}{E.~A. Hessels},
  \bibinfo{title}{A measurement of the atomic hydrogen Lamb shift and the
  proton charge radius}, \bibinfo{journal}{Science}
  \bibinfo{volume}{365}~(\bibinfo{number}{6457}) (\bibinfo{year}{2019})
  \bibinfo{pages}{1007--1012}.

\bibitem[{Xiong et~al.(2019)}]{Xiong:2019umf}
\bibinfo{author}{W.~Xiong}, et~al., \bibinfo{title}{{A small proton charge
  radius from an electron--proton scattering experiment}},
  \bibinfo{journal}{Nature} \bibinfo{volume}{575}~(\bibinfo{number}{7781})
  (\bibinfo{year}{2019}) \bibinfo{pages}{147--150}.

\bibitem[{Tiesinga et~al.(2021)Tiesinga, Mohr, Newell, and
  Taylor}]{Tiesinga:2021myr}
\bibinfo{author}{E.~Tiesinga}, \bibinfo{author}{P.~J. Mohr},
  \bibinfo{author}{D.~B. Newell}, \bibinfo{author}{B.~N. Taylor},
  \bibinfo{title}{{CODATA recommended values of the fundamental physical
  constants: 2018*}}, \bibinfo{journal}{Rev. Mod. Phys.}
  \bibinfo{volume}{93}~(\bibinfo{number}{2}) (\bibinfo{year}{2021})
  \bibinfo{pages}{025010}.

\bibitem[{Lin et~al.(2021{\natexlab{a}})Lin, Hammer, and
  Mei{\ss}ner}]{Lin:2021umk}
\bibinfo{author}{Y.-H. Lin}, \bibinfo{author}{H.-W. Hammer},
  \bibinfo{author}{U.-G. Mei{\ss}ner}, \bibinfo{title}{{High-precision
  determination of the electric and magnetic radius of the proton}},
  \bibinfo{journal}{Phys. Lett. B} \bibinfo{volume}{816}
  (\bibinfo{year}{2021}{\natexlab{a}}) \bibinfo{pages}{136254}.

\bibitem[{Cui et~al.(2021{\natexlab{a}})Cui, Binosi, Roberts, and
  Schmidt}]{Cui:2021vgm}
\bibinfo{author}{Z.-F. Cui}, \bibinfo{author}{D.~Binosi},
  \bibinfo{author}{C.~D. Roberts}, \bibinfo{author}{S.~M. Schmidt},
  \bibinfo{title}{{Fresh extraction of the proton charge radius from electron
  scattering}}, \bibinfo{journal}{Phys. Rev. Lett.}
  \bibinfo{volume}{127}~(\bibinfo{number}{9})
  (\bibinfo{year}{2021}{\natexlab{a}}) \bibinfo{pages}{092001}.

\bibitem[{Gilman et~al.(2013)}]{Gilman:2013eiv}
\bibinfo{author}{R.~Gilman}, et~al., \bibinfo{title}{{\emph{Studying the Proton
  ``Radius'' Puzzle with $\mu p$ Elastic Scattering} -- arXiv:1303.2160
  [hep-ex]$\!$}} .

\bibitem[{Cline et~al.(2021)Cline, Bernauer, Downie, and
  Gilman}]{Cline:2021ehf}
\bibinfo{author}{E.~Cline}, \bibinfo{author}{J.~Bernauer},
  \bibinfo{author}{E.~J. Downie}, \bibinfo{author}{R.~Gilman},
  \bibinfo{title}{{MUSE: The MUon Scattering Experiment}},
  \bibinfo{journal}{SciPost Phys. Proc.} \bibinfo{volume}{5}
  (\bibinfo{year}{2021}) \bibinfo{pages}{023}.

\bibitem[{Nakamura et~al.(2010)}]{Nakamura:2010zzi}
\bibinfo{author}{K.~Nakamura}, et~al., \bibinfo{title}{{Review of particle
  physics}}, \bibinfo{journal}{J. Phys. G} \bibinfo{volume}{37}
  (\bibinfo{year}{2010}) \bibinfo{pages}{075021}.

\bibitem[{Gao and Vanderhaeghen(2022)}]{Gao:2021sml}
\bibinfo{author}{H.~Gao}, \bibinfo{author}{M.~Vanderhaeghen},
  \bibinfo{title}{{The proton charge radius}}, \bibinfo{journal}{Rev. Mod.
  Phys.} \bibinfo{volume}{94}~(\bibinfo{number}{1}) (\bibinfo{year}{2022})
  \bibinfo{pages}{015002}.

\bibitem[{Durr et~al.(2008)}]{Durr:2008zz}
\bibinfo{author}{S.~Durr}, et~al., \bibinfo{title}{{Ab-Initio Determination of
  Light Hadron Masses}}, \bibinfo{journal}{Science} \bibinfo{volume}{322}
  (\bibinfo{year}{2008}) \bibinfo{pages}{1224--1227}.

\bibitem[{Qin and Roberts(2020)}]{Qin:2020rad}
\bibinfo{author}{S.-X. Qin}, \bibinfo{author}{C.~D. Roberts},
  \bibinfo{title}{{Impressions of the Continuum Bound State Problem in QCD}},
  \bibinfo{journal}{Chin. Phys. Lett.}
  \bibinfo{volume}{37}~(\bibinfo{number}{12}) (\bibinfo{year}{2020})
  \bibinfo{pages}{121201}.

\bibitem[{Carlson(2015)}]{Carlson:2015jba}
\bibinfo{author}{C.~E. Carlson}, \bibinfo{title}{{The Proton Radius Puzzle}},
  \bibinfo{journal}{Prog. Part. Nucl. Phys.} \bibinfo{volume}{82}
  (\bibinfo{year}{2015}) \bibinfo{pages}{59--77}.

\bibitem[{Kraus et~al.(2014)Kraus, Mesick, White, Gilman, and
  Strauch}]{Kraus:2014qua}
\bibinfo{author}{E.~Kraus}, \bibinfo{author}{K.~E. Mesick},
  \bibinfo{author}{A.~White}, \bibinfo{author}{R.~Gilman},
  \bibinfo{author}{S.~Strauch}, \bibinfo{title}{{Polynomial fits and the proton
  radius puzzle}}, \bibinfo{journal}{Phys. Rev. C} \bibinfo{volume}{90}
  (\bibinfo{year}{2014}) \bibinfo{pages}{045206}.

\bibitem[{Lorenz and Mei{\ss}ner(2014)}]{Lorenz:2014vha}
\bibinfo{author}{I.~T. Lorenz}, \bibinfo{author}{U.-G. Mei{\ss}ner},
  \bibinfo{title}{{Reduction of the proton radius discrepancy by $3\sigma$}},
  \bibinfo{journal}{Phys. Lett. B} \bibinfo{volume}{737} (\bibinfo{year}{2014})
  \bibinfo{pages}{57--59}.

\bibitem[{Griffioen et~al.(2016)Griffioen, Carlson, and
  Maddox}]{Griffioen:2015hta}
\bibinfo{author}{K.~Griffioen}, \bibinfo{author}{C.~Carlson},
  \bibinfo{author}{S.~Maddox}, \bibinfo{title}{{Consistency of electron
  scattering data with a small proton radius}}, \bibinfo{journal}{Phys. Rev. C}
  \bibinfo{volume}{93} (\bibinfo{year}{2016}) \bibinfo{pages}{065207}.

\bibitem[{Higinbotham et~al.(2016)Higinbotham, Kabir, Lin, Meekins, Norum, and
  Sawatzky}]{Higinbotham:2015rja}
\bibinfo{author}{D.~W. Higinbotham}, \bibinfo{author}{A.~A. Kabir},
  \bibinfo{author}{V.~Lin}, \bibinfo{author}{D.~Meekins},
  \bibinfo{author}{B.~Norum}, \bibinfo{author}{B.~Sawatzky},
  \bibinfo{title}{{Proton radius from electron scattering data}},
  \bibinfo{journal}{Phys. Rev. C} \bibinfo{volume}{93} (\bibinfo{year}{2016})
  \bibinfo{pages}{055207}.

\bibitem[{Hayward and Griffioen(2020)}]{Hayward:2018qij}
\bibinfo{author}{T.~B. Hayward}, \bibinfo{author}{K.~A. Griffioen},
  \bibinfo{title}{{Evaluation of low-$Q^2$ fits to $ep$ and $ed$ elastic
  scattering data}}, \bibinfo{journal}{Nucl. Phys. A} \bibinfo{volume}{999}
  (\bibinfo{year}{2020}) \bibinfo{pages}{121767}.

\bibitem[{Zhou et~al.(2019)Zhou, Giulani, Piekarewicz, Bhattacharya, and
  Pati}]{Zhou:2018bon}
\bibinfo{author}{S.~Zhou}, \bibinfo{author}{P.~Giulani},
  \bibinfo{author}{J.~Piekarewicz}, \bibinfo{author}{A.~Bhattacharya},
  \bibinfo{author}{D.~Pati}, \bibinfo{title}{{Reexamining the proton-radius
  problem using constrained Gaussian processes}}, \bibinfo{journal}{Phys. Rev.
  C} \bibinfo{volume}{99} (\bibinfo{year}{2019}) \bibinfo{pages}{055202}.

\bibitem[{Alarc{\'o}n et~al.(2019)Alarc{\'o}n, Higinbotham, Weiss, and
  Ye}]{Alarcon:2018zbz}
\bibinfo{author}{J.~M. Alarc{\'o}n}, \bibinfo{author}{D.~W. Higinbotham},
  \bibinfo{author}{C.~Weiss}, \bibinfo{author}{Z.~Ye}, \bibinfo{title}{{Proton
  charge radius extraction from electron scattering data using dispersively
  improved chiral effective field theory}}, \bibinfo{journal}{Phys. Rev. C}
  \bibinfo{volume}{99} (\bibinfo{year}{2019}) \bibinfo{pages}{044303}.

\bibitem[{Barcus et~al.(2020)Barcus, Higinbotham, and
  McClellan}]{Higinbotham:2019jzd}
\bibinfo{author}{S.~K. Barcus}, \bibinfo{author}{D.~W. Higinbotham},
  \bibinfo{author}{R.~E. McClellan}, \bibinfo{title}{{How Analytic Choices Can
  Affect the Extraction of Electromagnetic Form Factors from Elastic Electron
  Scattering Cross Section Data}}, \bibinfo{journal}{Phys. Rev. C}
  \bibinfo{volume}{102}~(\bibinfo{number}{1}) (\bibinfo{year}{2020})
  \bibinfo{pages}{015205}.

\bibitem[{Hammer and Mei{\ss}ner(2020)}]{Hammer:2019uab}
\bibinfo{author}{H.-W. Hammer}, \bibinfo{author}{U.-G. Mei{\ss}ner},
  \bibinfo{title}{{The proton radius: From a puzzle to precision}},
  \bibinfo{journal}{Sci. Bull.} \bibinfo{volume}{65} (\bibinfo{year}{2020})
  \bibinfo{pages}{257--258}.

\bibitem[{Yan et~al.(2018)Yan, Higinbotham, Dutta, Gao, Gasparian, Khandaker,
  Liyanage, Pasyuk, Peng, and Xiong}]{Yan:2018bez}
\bibinfo{author}{X.~Yan}, \bibinfo{author}{D.~W. Higinbotham},
  \bibinfo{author}{D.~Dutta}, \bibinfo{author}{H.~Gao},
  \bibinfo{author}{A.~Gasparian}, \bibinfo{author}{M.~A. Khandaker},
  \bibinfo{author}{N.~Liyanage}, \bibinfo{author}{E.~Pasyuk},
  \bibinfo{author}{C.~Peng}, \bibinfo{author}{W.~Xiong},
  \bibinfo{title}{{Robust extraction of the proton charge radius from
  electron-proton scattering data}}, \bibinfo{journal}{Phys. Rev. C}
  \bibinfo{volume}{98} (\bibinfo{year}{2018}) \bibinfo{pages}{025204}.

\bibitem[{Reinsch(1967)}]{Reinsch:1967aa}
\bibinfo{author}{C.~H. Reinsch}, \bibinfo{title}{Smoothing by spline
  functions}, \bibinfo{journal}{Numer. Math.} \bibinfo{volume}{10}
  (\bibinfo{year}{1967}) \bibinfo{pages}{177--183}.

\bibitem[{Borkowski et~al.(1975)Borkowski, Simon, Walther, and
  Wendling}]{Borkowski:1975ume}
\bibinfo{author}{F.~Borkowski}, \bibinfo{author}{G.~G. Simon},
  \bibinfo{author}{V.~H. Walther}, \bibinfo{author}{R.~D. Wendling},
  \bibinfo{title}{{On the determination of the proton RMS-radius from electron
  scattering data}}, \bibinfo{journal}{Z. Phys. A}
  \bibinfo{volume}{275}~(\bibinfo{number}{1}) (\bibinfo{year}{1975})
  \bibinfo{pages}{29--31}.

\bibitem[{Kelly(2004)}]{Kelly:2004hm}
\bibinfo{author}{J.~J. Kelly}, \bibinfo{title}{{Simple parametrization of
  nucleon form factors}}, \bibinfo{journal}{Phys. Rev. C} \bibinfo{volume}{70}
  (\bibinfo{year}{2004}) \bibinfo{pages}{068202}.

\bibitem[{Arrington(2004)}]{Arrington:2003qk}
\bibinfo{author}{J.~Arrington}, \bibinfo{title}{{Implications of the
  discrepancy between proton form-factor measurements}},
  \bibinfo{journal}{Phys. Rev. C} \bibinfo{volume}{69} (\bibinfo{year}{2004})
  \bibinfo{pages}{022201}.

\bibitem[{Arrington and Sick(2007)}]{Arrington:2006hm}
\bibinfo{author}{J.~Arrington}, \bibinfo{author}{I.~Sick},
  \bibinfo{title}{{Precise determination of low-Q nucleon electromagnetic form
  factors and their impact on parity-violating e-p elastic scattering}},
  \bibinfo{journal}{Phys. Rev. C} \bibinfo{volume}{76} (\bibinfo{year}{2007})
  \bibinfo{pages}{035201}.

\bibitem[{Bernauer et~al.(2014{\natexlab{a}})}]{Bernauer:2013tpr}
\bibinfo{author}{J.~C. Bernauer}, et~al., \bibinfo{title}{{Electric and
  magnetic form factors of the proton}}, \bibinfo{journal}{Phys. Rev. C}
  \bibinfo{volume}{90}~(\bibinfo{number}{1})
  (\bibinfo{year}{2014}{\natexlab{a}}) \bibinfo{pages}{015206}.

\bibitem[{Ye et~al.(2018)Ye, Arrington, Hill, and Lee}]{Ye:2017gyb}
\bibinfo{author}{Z.~Ye}, \bibinfo{author}{J.~Arrington}, \bibinfo{author}{R.~J.
  Hill}, \bibinfo{author}{G.~Lee}, \bibinfo{title}{{Proton and Neutron
  Electromagnetic Form Factors and Uncertainties}}, \bibinfo{journal}{Phys.
  Lett. B} \bibinfo{volume}{777} (\bibinfo{year}{2018}) \bibinfo{pages}{8--15}.

\bibitem[{Alarc\'on and Weiss(2017)}]{Alarcon:2017ivh}
\bibinfo{author}{J.~M. Alarc\'on}, \bibinfo{author}{C.~Weiss},
  \bibinfo{title}{{Nucleon form factors in dispersively improved chiral
  effective field theory: Scalar form factor}}, \bibinfo{journal}{Phys. Rev. C}
  \bibinfo{volume}{96}~(\bibinfo{number}{5}) (\bibinfo{year}{2017})
  \bibinfo{pages}{055206}.

\bibitem[{{A.~Gasparian, H.~Gao, D.~Dutta, N.~Liyanage, E.~Pasyuk, et
  al.}(2017)}]{JlabDRad}
\bibinfo{author}{{A.~Gasparian, H.~Gao, D.~Dutta, N.~Liyanage, E.~Pasyuk, et
  al.}}, \bibinfo{title}{Precision deuteron charge radius measurement with
  elastic electron-deuteron scattering -- PRad Collaboration}
  \bibinfo{note}{\mbox{ }Jefferson Lab proposal PR12-17-009}.

\bibitem[{Grinin et~al.(2020)Grinin, Matveev, Yost, Maisenbacher, Wirthl, Pohl,
  H{\"a}nsch, and Udem}]{Grinin1061}
\bibinfo{author}{A.~Grinin}, \bibinfo{author}{A.~Matveev},
  \bibinfo{author}{D.~C. Yost}, \bibinfo{author}{L.~Maisenbacher},
  \bibinfo{author}{V.~Wirthl}, \bibinfo{author}{R.~Pohl},
  \bibinfo{author}{T.~W. H{\"a}nsch}, \bibinfo{author}{T.~Udem},
  \bibinfo{title}{Two-photon frequency comb spectroscopy of atomic hydrogen},
  \bibinfo{journal}{Science} \bibinfo{volume}{370}~(\bibinfo{number}{6520})
  (\bibinfo{year}{2020}) \bibinfo{pages}{1061--1066}.

\bibitem[{Pohl et~al.(2016{\natexlab{a}})}]{Pohl1:2016xoo}
\bibinfo{author}{R.~Pohl}, et~al., \bibinfo{title}{{Laser spectroscopy of
  muonic deuterium}}, \bibinfo{journal}{Science}
  \bibinfo{volume}{353}~(\bibinfo{number}{6300})
  (\bibinfo{year}{2016}{\natexlab{a}}) \bibinfo{pages}{669--673}.

\bibitem[{Sachs(1962)}]{Sachs:1962zzc}
\bibinfo{author}{R.~Sachs}, \bibinfo{title}{{High-Energy Behavior of Nucleon
  Electromagnetic Form Factors}}, \bibinfo{journal}{Phys. Rev.}
  \bibinfo{volume}{126} (\bibinfo{year}{1962}) \bibinfo{pages}{2256--2260}.

\bibitem[{Miller(2010)}]{Miller:2010nz}
\bibinfo{author}{G.~A. Miller}, \bibinfo{title}{{Transverse Charge Densities}},
  \bibinfo{journal}{Ann. Rev. Nucl. Part. Sci.} \bibinfo{volume}{60}
  (\bibinfo{year}{2010}) \bibinfo{pages}{1--25}.

\bibitem[{Aoyama et~al.(2012{\natexlab{a}})Aoyama, Hayakawa, Kinoshita, and
  Nio}]{Aoyama:2012wj}
\bibinfo{author}{T.~Aoyama}, \bibinfo{author}{M.~Hayakawa},
  \bibinfo{author}{T.~Kinoshita}, \bibinfo{author}{M.~Nio},
  \bibinfo{title}{{Tenth-Order QED Contribution to the Electron g-2 and an
  Improved Value of the Fine Structure Constant}}, \bibinfo{journal}{Phys. Rev.
  Lett.} \bibinfo{volume}{109} (\bibinfo{year}{2012}{\natexlab{a}})
  \bibinfo{pages}{111807}.

\bibitem[{Aoyama et~al.(2012{\natexlab{b}})Aoyama, Hayakawa, Kinoshita, and
  Nio}]{Aoyama:2012wk}
\bibinfo{author}{T.~Aoyama}, \bibinfo{author}{M.~Hayakawa},
  \bibinfo{author}{T.~Kinoshita}, \bibinfo{author}{M.~Nio},
  \bibinfo{title}{{Complete Tenth-Order QED Contribution to the Muon g-2}},
  \bibinfo{journal}{Phys. Rev. Lett.} \bibinfo{volume}{109}
  (\bibinfo{year}{2012}{\natexlab{b}}) \bibinfo{pages}{111808}.

\bibitem[{Jones et~al.(2000)}]{Jones:1999rz}
\bibinfo{author}{M.~K. Jones}, et~al., \bibinfo{title}{{$G_{E_p}/G_{M_p}$ ratio
  by polarization transfer in $\vec{e} p \to e\vec{p}$}},
  \bibinfo{journal}{Phys. Rev. Lett.} \bibinfo{volume}{84}
  (\bibinfo{year}{2000}) \bibinfo{pages}{1398--1402}.

\bibitem[{Foldy(1958)}]{Foldy:1958zz}
\bibinfo{author}{L.~L. Foldy}, \bibinfo{title}{{Neutron-Electron Interaction}},
  \bibinfo{journal}{Rev. Mod. Phys.} \bibinfo{volume}{30}
  (\bibinfo{year}{1958}) \bibinfo{pages}{471--481}.

\bibitem[{Bernauer et~al.(2014{\natexlab{b}})}]{A1:2013fsc}
\bibinfo{author}{J.~C. Bernauer}, et~al., \bibinfo{title}{{Electric and
  magnetic form factors of the proton}}, \bibinfo{journal}{Phys. Rev. C}
  \bibinfo{volume}{90}~(\bibinfo{number}{1})
  (\bibinfo{year}{2014}{\natexlab{b}}) \bibinfo{pages}{015206}.

\bibitem[{Lee et~al.(2015)Lee, Arrington, and Hill}]{Lee:2015jqa}
\bibinfo{author}{G.~Lee}, \bibinfo{author}{J.~R. Arrington},
  \bibinfo{author}{R.~J. Hill}, \bibinfo{title}{{Extraction of the proton
  radius from electron-proton scattering data}}, \bibinfo{journal}{Phys. Rev.
  D} \bibinfo{volume}{92}~(\bibinfo{number}{1}) (\bibinfo{year}{2015})
  \bibinfo{pages}{013013}.

\bibitem[{Alarc\'on et~al.(2020)Alarc\'on, Higinbotham, and
  Weiss}]{Alarcon:2020kcz}
\bibinfo{author}{J.~M. Alarc\'on}, \bibinfo{author}{D.~W. Higinbotham},
  \bibinfo{author}{C.~Weiss}, \bibinfo{title}{{Precise determination of the
  proton magnetic radius from electron scattering data}},
  \bibinfo{journal}{Phys. Rev. C} \bibinfo{volume}{102}~(\bibinfo{number}{3})
  (\bibinfo{year}{2020}) \bibinfo{pages}{035203}.

\bibitem[{Lin et~al.(2021{\natexlab{b}})Lin, Hammer, and
  Mei\ss{}ner}]{Lin:2021umz}
\bibinfo{author}{Y.-H. Lin}, \bibinfo{author}{H.-W. Hammer},
  \bibinfo{author}{U.-G. Mei\ss{}ner}, \bibinfo{title}{{Dispersion-theoretical
  analysis of the electromagnetic form factors of the nucleon: Past, present
  and future}}, \bibinfo{journal}{Eur. Phys. J. A} \bibinfo{volume}{57}
  (\bibinfo{year}{2021}{\natexlab{b}}) \bibinfo{pages}{255}.

\bibitem[{Djukanovic et~al.(2021)Djukanovic, Harris, von Hippel, Junnarkar,
  Meyer, Mohler, Ottnad, Schulz, Wilhelm, and Wittig}]{Djukanovic:2021cgp}
\bibinfo{author}{D.~Djukanovic}, \bibinfo{author}{T.~Harris},
  \bibinfo{author}{G.~von Hippel}, \bibinfo{author}{P.~M. Junnarkar},
  \bibinfo{author}{H.~B. Meyer}, \bibinfo{author}{D.~Mohler},
  \bibinfo{author}{K.~Ottnad}, \bibinfo{author}{T.~Schulz},
  \bibinfo{author}{J.~Wilhelm}, \bibinfo{author}{H.~Wittig},
  \bibinfo{title}{{Isovector electromagnetic form factors of the nucleon from
  lattice QCD and the proton radius puzzle}}, \bibinfo{journal}{Phys. Rev. D}
  \bibinfo{volume}{103}~(\bibinfo{number}{9}) (\bibinfo{year}{2021})
  \bibinfo{pages}{094522}.

\bibitem[{Cui et~al.(2021{\natexlab{b}})Cui, Binosi, Roberts, and
  Schmidt}]{Cui:2021skn}
\bibinfo{author}{Z.-F. Cui}, \bibinfo{author}{D.~Binosi},
  \bibinfo{author}{C.~D. Roberts}, \bibinfo{author}{S.~M. Schmidt},
  \bibinfo{title}{{Pauli radius of the proton}}, \bibinfo{journal}{Chin. Phys.
  Lett. \emph{Express}} \bibinfo{volume}{38}~(\bibinfo{number}{12})
  (\bibinfo{year}{2021}{\natexlab{b}}) \bibinfo{pages}{121401}.

\bibitem[{Rosenbluth(1950)}]{Rosenbluth:1950yq}
\bibinfo{author}{M.~Rosenbluth}, \bibinfo{title}{{High Energy Elastic
  Scattering of Electrons on Protons}}, \bibinfo{journal}{Phys. Rev.}
  \bibinfo{volume}{79} (\bibinfo{year}{1950}) \bibinfo{pages}{615--619}.

\bibitem[{Brodsky et~al.(1998)Brodsky, Pauli, and Pinsky}]{Brodsky:1997de}
\bibinfo{author}{S.~J. Brodsky}, \bibinfo{author}{H.-C. Pauli},
  \bibinfo{author}{S.~S. Pinsky}, \bibinfo{title}{{Quantum chromodynamics and
  other field theories on the light cone}}, \bibinfo{journal}{Phys. Rept.}
  \bibinfo{volume}{301} (\bibinfo{year}{1998}) \bibinfo{pages}{299--486}.

\bibitem[{Brodsky et~al.(2022)Brodsky, Deur, and Roberts}]{Brodsky:2022fqy}
\bibinfo{author}{S.~J. Brodsky}, \bibinfo{author}{A.~Deur},
  \bibinfo{author}{C.~D. Roberts}, \bibinfo{title}{{Artificial Dynamical
  Effects in Quantum Field Theory}}, \bibinfo{journal}{Nature Reviews Physics}
  (\bibinfo{year}{2022}) \bibinfo{pages}{May}.

\bibitem[{Miller et~al.(2008)Miller, Piasetzky, and Ron}]{Miller:2007kt}
\bibinfo{author}{G.~A. Miller}, \bibinfo{author}{E.~Piasetzky},
  \bibinfo{author}{G.~Ron}, \bibinfo{title}{{Proton Electromagnetic Form Factor
  Ratios at Low $Q^2$}}, \bibinfo{journal}{Phys. Rev. Lett.}
  \bibinfo{volume}{101} (\bibinfo{year}{2008}) \bibinfo{pages}{082002}.

\bibitem[{Punjabi et~al.(2015)Punjabi, Perdrisat, Jones, Brash, and
  Carlson}]{Punjabi:2015bba}
\bibinfo{author}{V.~Punjabi}, \bibinfo{author}{C.~F. Perdrisat},
  \bibinfo{author}{M.~K. Jones}, \bibinfo{author}{E.~J. Brash},
  \bibinfo{author}{C.~E. Carlson}, \bibinfo{title}{{The Structure of the
  Nucleon: Elastic Electromagnetic Form Factors}}, \bibinfo{journal}{Eur. Phys.
  J. A} \bibinfo{volume}{51} (\bibinfo{year}{2015}) \bibinfo{pages}{79}.

\bibitem[{Cloet and Roberts(2014)}]{Cloet:2013jya}
\bibinfo{author}{I.~C. Cloet}, \bibinfo{author}{C.~D. Roberts},
  \bibinfo{title}{{Explanation and Prediction of Observables using Continuum
  Strong QCD}}, \bibinfo{journal}{Prog. Part. Nucl. Phys.} \bibinfo{volume}{77}
  (\bibinfo{year}{2014}) \bibinfo{pages}{1--69}.

\bibitem[{Barabanov et~al.(2021)}]{Barabanov:2020jvn}
\bibinfo{author}{M.~Y. Barabanov}, et~al., \bibinfo{title}{{Diquark
  Correlations in Hadron Physics: Origin, Impact and Evidence}},
  \bibinfo{journal}{Prog. Part. Nucl. Phys.} \bibinfo{volume}{116}
  (\bibinfo{year}{2021}) \bibinfo{pages}{103835}.

\bibitem[{Sufian et~al.(2017)Sufian, de~T\'eramond, Brodsky, Deur, and
  Dosch}]{Sufian:2016hwn}
\bibinfo{author}{R.~S. Sufian}, \bibinfo{author}{G.~F. de~T\'eramond},
  \bibinfo{author}{S.~J. Brodsky}, \bibinfo{author}{A.~Deur},
  \bibinfo{author}{H.~G. Dosch}, \bibinfo{title}{{Analysis of nucleon
  electromagnetic form factors from light-front holographic QCD : The spacelike
  region}}, \bibinfo{journal}{Phys. Rev. D}
  \bibinfo{volume}{95}~(\bibinfo{number}{1}) (\bibinfo{year}{2017})
  \bibinfo{pages}{014011}.

\bibitem[{Xu et~al.(2019)Xu, Binosi, Cui, Li, Roberts, Xu, and
  Zong}]{Xu:2019ilh}
\bibinfo{author}{Y.-Z. Xu}, \bibinfo{author}{D.~Binosi}, \bibinfo{author}{Z.-F.
  Cui}, \bibinfo{author}{B.-L. Li}, \bibinfo{author}{C.~D. Roberts},
  \bibinfo{author}{S.-S. Xu}, \bibinfo{author}{H.-S. Zong},
  \bibinfo{title}{{Elastic electromagnetic form factors of vector mesons}},
  \bibinfo{journal}{Phys. Rev. D} \bibinfo{volume}{100} (\bibinfo{year}{2019})
  \bibinfo{pages}{114038}.

\bibitem[{Mondal et~al.(2020)Mondal, Xu, Lan, Zhao, Li, Chakrabarti, and
  Vary}]{Mondal:2019jdg}
\bibinfo{author}{C.~Mondal}, \bibinfo{author}{S.~Xu}, \bibinfo{author}{J.~Lan},
  \bibinfo{author}{X.~Zhao}, \bibinfo{author}{Y.~Li},
  \bibinfo{author}{D.~Chakrabarti}, \bibinfo{author}{J.~P. Vary},
  \bibinfo{title}{{Proton structure from a light-front Hamiltonian}},
  \bibinfo{journal}{Phys. Rev. D} \bibinfo{volume}{102}~(\bibinfo{number}{1})
  (\bibinfo{year}{2020}) \bibinfo{pages}{016008}.

\bibitem[{Cui et~al.(2020)Cui, Chen, Binosi, de~Soto, Roberts,
  Rodr{\'{\i}}guez-Quintero, Schmidt, and Segovia}]{Cui:2020rmu}
\bibinfo{author}{Z.-F. Cui}, \bibinfo{author}{C.~Chen},
  \bibinfo{author}{D.~Binosi}, \bibinfo{author}{F.~de~Soto},
  \bibinfo{author}{C.~D. Roberts},
  \bibinfo{author}{J.~Rodr{\'{\i}}guez-Quintero}, \bibinfo{author}{S.~M.
  Schmidt}, \bibinfo{author}{J.~Segovia}, \bibinfo{title}{{Nucleon elastic form
  factors at accessible large spacelike momenta}}, \bibinfo{journal}{Phys. Rev.
  D} \bibinfo{volume}{102} (\bibinfo{year}{2020}) \bibinfo{pages}{014043}.

\bibitem[{Lane(1974)}]{Lane:1974he}
\bibinfo{author}{K.~D. Lane}, \bibinfo{title}{{Asymptotic Freedom and Goldstone
  Realization of Chiral Symmetry}}, \bibinfo{journal}{Phys. Rev. D}
  \bibinfo{volume}{10} (\bibinfo{year}{1974}) \bibinfo{pages}{2605}.

\bibitem[{Politzer(1976)}]{Politzer:1976tv}
\bibinfo{author}{H.~D. Politzer}, \bibinfo{title}{{Effective Quark Masses in
  the Chiral Limit}}, \bibinfo{journal}{Nucl. Phys. B} \bibinfo{volume}{117}
  (\bibinfo{year}{1976}) \bibinfo{pages}{397}.

\bibitem[{Delbourgo and Scadron(1979)}]{Delbourgo:1979me}
\bibinfo{author}{R.~Delbourgo}, \bibinfo{author}{M.~D. Scadron},
  \bibinfo{title}{{Proof of the Nambu-Goldstone realization for vector gluon
  quark theories}}, \bibinfo{journal}{J. Phys. G} \bibinfo{volume}{5}
  (\bibinfo{year}{1979}) \bibinfo{pages}{1621}.

\bibitem[{Maris et~al.(1998)Maris, Roberts, and Tandy}]{Maris:1997hd}
\bibinfo{author}{P.~Maris}, \bibinfo{author}{C.~D. Roberts},
  \bibinfo{author}{P.~C. Tandy}, \bibinfo{title}{{Pion mass and decay
  constant}}, \bibinfo{journal}{Phys. Lett. B} \bibinfo{volume}{420}
  (\bibinfo{year}{1998}) \bibinfo{pages}{267--273}.

\bibitem[{Brodsky et~al.(2012)Brodsky, Roberts, Shrock, and
  Tandy}]{Brodsky:2012ku}
\bibinfo{author}{S.~J. Brodsky}, \bibinfo{author}{C.~D. Roberts},
  \bibinfo{author}{R.~Shrock}, \bibinfo{author}{P.~C. Tandy},
  \bibinfo{title}{{Confinement contains condensates}}, \bibinfo{journal}{Phys.
  Rev. C} \bibinfo{volume}{85} (\bibinfo{year}{2012}) \bibinfo{pages}{065202}.

\bibitem[{Yukawa(1935)}]{Yukawa:1935xg}
\bibinfo{author}{H.~Yukawa}, \bibinfo{title}{{On the interaction of elementary
  particles}}, \bibinfo{journal}{Proc. Phys. Math. Soc. Jap.}
  \bibinfo{volume}{17} (\bibinfo{year}{1935}) \bibinfo{pages}{48--57}.

\bibitem[{Lattes et~al.(1947)Lattes, Muirhead, Occhialini, and
  Powell}]{Lattes:1947mw}
\bibinfo{author}{C.~M.~G. Lattes}, \bibinfo{author}{H.~Muirhead},
  \bibinfo{author}{G.~P.~S. Occhialini}, \bibinfo{author}{C.~F. Powell},
  \bibinfo{title}{{Processes involving charged mesons}},
  \bibinfo{journal}{Nature} \bibinfo{volume}{159} (\bibinfo{year}{1947})
  \bibinfo{pages}{694--697}.

\bibitem[{Chang and Roberts(2021)}]{Chang:2021utv}
\bibinfo{author}{L.~Chang}, \bibinfo{author}{C.~D. Roberts},
  \bibinfo{title}{{Regarding the distribution of glue in the pion}},
  \bibinfo{journal}{Chin. Phys. Lett.}
  \bibinfo{volume}{38}~(\bibinfo{number}{8}) (\bibinfo{year}{2021})
  \bibinfo{pages}{081101}.

\bibitem[{Chang et~al.(2022)Chang, Gao, and Roberts}]{Chang:2022jri}
\bibinfo{author}{L.~Chang}, \bibinfo{author}{F.~Gao}, \bibinfo{author}{C.~D.
  Roberts}, \bibinfo{title}{{Parton distributions of light quarks and
  antiquarks in the proton}}, \bibinfo{journal}{Phys. Lett. B}
  \bibinfo{volume}{829} (\bibinfo{year}{2022}) \bibinfo{pages}{137078}.

\bibitem[{Lu et~al.(2022)Lu, Chang, Raya, Roberts, and
  Rodr\'\i{}guez-Quintero}]{Lu:2022cjx}
\bibinfo{author}{Y.~Lu}, \bibinfo{author}{L.~Chang}, \bibinfo{author}{K.~Raya},
  \bibinfo{author}{C.~D. Roberts},
  \bibinfo{author}{J.~Rodr\'\i{}guez-Quintero}, \bibinfo{title}{{Proton and
  pion distribution functions in counterpoint}}, \bibinfo{journal}{Phys. Lett.
  B} \bibinfo{volume}{830} (\bibinfo{year}{2022}) \bibinfo{pages}{137130}.

\bibitem[{Chen et~al.(2018)Chen, Ding, Chang, and Roberts}]{Chen:2018rwz}
\bibinfo{author}{M.~Chen}, \bibinfo{author}{M.~Ding},
  \bibinfo{author}{L.~Chang}, \bibinfo{author}{C.~D. Roberts},
  \bibinfo{title}{{Mass-dependence of pseudoscalar meson elastic form
  factors}}, \bibinfo{journal}{Phys. Rev. D} \bibinfo{volume}{98}
  (\bibinfo{year}{2018}) \bibinfo{pages}{091505(R)}.

\bibitem[{Amendolia et~al.(1984)}]{Amendolia:1984nz}
\bibinfo{author}{S.~R. Amendolia}, et~al., \bibinfo{title}{{A Measurement of
  the Pion Charge Radius}}, \bibinfo{journal}{Phys. Lett. B}
  \bibinfo{volume}{146} (\bibinfo{year}{1984}) \bibinfo{pages}{116}.

\bibitem[{Amendolia et~al.(1986{\natexlab{a}})}]{Amendolia:1986wj}
\bibinfo{author}{S.~R. Amendolia}, et~al., \bibinfo{title}{{A Measurement of
  the Space - Like Pion Electromagnetic Form-Factor}}, \bibinfo{journal}{Nucl.
  Phys. B} \bibinfo{volume}{277} (\bibinfo{year}{1986}{\natexlab{a}})
  \bibinfo{pages}{168}.

\bibitem[{Dally et~al.(1982)}]{Dally:1982zk}
\bibinfo{author}{E.~B. Dally}, et~al., \bibinfo{title}{{Elastic Scattering
  Measurement of the Negative Pion Radius}}, \bibinfo{journal}{Phys. Rev.
  Lett.} \bibinfo{volume}{48} (\bibinfo{year}{1982}) \bibinfo{pages}{375--378}.

\bibitem[{Gough~Eschrich et~al.(2001)}]{GoughEschrich:2001ji}
\bibinfo{author}{I.~M. Gough~Eschrich}, et~al., \bibinfo{title}{{Measurement of
  the Sigma- Charge Radius by Sigma- Electron Elastic Scattering}},
  \bibinfo{journal}{Phys. Lett. B} \bibinfo{volume}{522} (\bibinfo{year}{2001})
  \bibinfo{pages}{233--239}.

\bibitem[{Ananthanarayan et~al.(2017)Ananthanarayan, Caprini, and
  Das}]{Ananthanarayan:2017efc}
\bibinfo{author}{B.~Ananthanarayan}, \bibinfo{author}{I.~Caprini},
  \bibinfo{author}{D.~Das}, \bibinfo{title}{{Electromagnetic charge radius of
  the pion at high precision}}, \bibinfo{journal}{Phys. Rev. Lett.}
  \bibinfo{volume}{119}~(\bibinfo{number}{13}) (\bibinfo{year}{2017})
  \bibinfo{pages}{132002}.

\bibitem[{Colangelo et~al.(2019)Colangelo, Hoferichter, and
  Stoffer}]{Colangelo:2018mtw}
\bibinfo{author}{G.~Colangelo}, \bibinfo{author}{M.~Hoferichter},
  \bibinfo{author}{P.~Stoffer}, \bibinfo{title}{{Two-pion contribution to
  hadronic vacuum polarization}}, \bibinfo{journal}{JHEP} \bibinfo{volume}{02}
  (\bibinfo{year}{2019}) \bibinfo{pages}{006}.

\bibitem[{Tanabashi et~al.(2018)}]{Tanabashi:2018oca}
\bibinfo{author}{M.~Tanabashi}, et~al., \bibinfo{title}{{Review of Particle
  Physics}}, \bibinfo{journal}{Phys. Rev. D} \bibinfo{volume}{98}
  (\bibinfo{year}{2018}) \bibinfo{pages}{030001}.

\bibitem[{Cui et~al.(2021{\natexlab{c}})Cui, Binosi, Roberts, and
  Schmidt}]{Cui:2021aee}
\bibinfo{author}{Z.-F. Cui}, \bibinfo{author}{D.~Binosi},
  \bibinfo{author}{C.~D. Roberts}, \bibinfo{author}{S.~M. Schmidt},
  \bibinfo{title}{{Pion charge radius from pion+electron elastic scattering
  data}}, \bibinfo{journal}{Phys. Lett. B} \bibinfo{volume}{822}
  (\bibinfo{year}{2021}{\natexlab{c}}) \bibinfo{pages}{136631}.

\bibitem[{Dally et~al.(1980)Dally, Hauptman, Kubic, Stork, Watson
  et~al.}]{Dally:1980dj}
\bibinfo{author}{E.~Dally}, \bibinfo{author}{J.~Hauptman},
  \bibinfo{author}{J.~Kubic}, \bibinfo{author}{D.~Stork},
  \bibinfo{author}{A.~Watson}, et~al., \bibinfo{title}{{Direct Measurement of
  the Negative-Kaon Form Factor}}, \bibinfo{journal}{Phys. Rev. Lett.}
  \bibinfo{volume}{45} (\bibinfo{year}{1980}) \bibinfo{pages}{232--235}.

\bibitem[{Amendolia et~al.(1986{\natexlab{b}})}]{Amendolia:1986ui}
\bibinfo{author}{S.~Amendolia}, et~al., \bibinfo{title}{{A Measurement of the
  Kaon Charge Radius}}, \bibinfo{journal}{Phys. Lett. B} \bibinfo{volume}{178}
  (\bibinfo{year}{1986}{\natexlab{b}}) \bibinfo{pages}{435}.

\bibitem[{Adams et~al.(2021)Adams, Carleo, Lovato, and Rocco}]{Adams:2020aax}
\bibinfo{author}{C.~Adams}, \bibinfo{author}{G.~Carleo},
  \bibinfo{author}{A.~Lovato}, \bibinfo{author}{N.~Rocco},
  \bibinfo{title}{{Variational Monte Carlo Calculations of A\ensuremath{\leq}4
  Nuclei with an Artificial Neural-Network Correlator Ansatz}},
  \bibinfo{journal}{Phys. Rev. Lett.}
  \bibinfo{volume}{127}~(\bibinfo{number}{2}) (\bibinfo{year}{2021})
  \bibinfo{pages}{022502}.

\bibitem[{Pohl et~al.(2016{\natexlab{b}})}]{CREMA:2016idx}
\bibinfo{author}{R.~Pohl}, et~al., \bibinfo{title}{{Laser spectroscopy of
  muonic deuterium}}, \bibinfo{journal}{Science}
  \bibinfo{volume}{353}~(\bibinfo{number}{6300})
  (\bibinfo{year}{2016}{\natexlab{b}}) \bibinfo{pages}{669--673}.

\bibitem[{Pohl et~al.(2017)}]{Pohl:2016glp}
\bibinfo{author}{R.~Pohl}, et~al., \bibinfo{title}{{Deuteron charge radius and
  Rydberg constant from spectroscopy data in atomic deuterium}},
  \bibinfo{journal}{Metrologia} \bibinfo{volume}{54}~(\bibinfo{number}{2})
  (\bibinfo{year}{2017}) \bibinfo{pages}{L1}.

\bibitem[{Zhou et~al.(2021)}]{Zhou:2020cdt}
\bibinfo{author}{J.~Zhou}, et~al., \bibinfo{title}{{Advanced extraction of the
  deuteron charge radius from electron-deuteron scattering data}},
  \bibinfo{journal}{Phys. Rev. C} \bibinfo{volume}{103}~(\bibinfo{number}{2})
  (\bibinfo{year}{2021}) \bibinfo{pages}{024002}.

\bibitem[{Abbott et~al.(2000)}]{JLABt20:2000qyq}
\bibinfo{author}{D.~Abbott}, et~al., \bibinfo{title}{{Phenomenology of the
  deuteron electromagnetic form-factors}}, \bibinfo{journal}{Eur. Phys. J. A}
  \bibinfo{volume}{7} (\bibinfo{year}{2000}) \bibinfo{pages}{421--427}.

\bibitem[{Kobushkin and Syamtomov(1995)}]{Kobushkin:1994ed}
\bibinfo{author}{A.~P. Kobushkin}, \bibinfo{author}{A.~I. Syamtomov},
  \bibinfo{title}{{Deuteron electromagnetic form-factors in the transitional
  region between nucleon - meson and quark - gluon pictures}},
  \bibinfo{journal}{Phys. Atom. Nucl.} \bibinfo{volume}{58}
  (\bibinfo{year}{1995}) \bibinfo{pages}{1477--1482}.

\bibitem[{Parker and Higinbotham(2020)}]{asia_parker_2020_4074281}
\bibinfo{author}{A.~Parker}, \bibinfo{author}{D.~W. Higinbotham},
  \bibinfo{title}{Deuteron Form Factor Parameterization},
  \urlprefix\url{https://doi.org/10.5281/zenodo.4074281}, \bibinfo{year}{2020}.

\bibitem[{Sick(1974)}]{Sick:1974suq}
\bibinfo{author}{I.~Sick}, \bibinfo{title}{{Model-independent nuclear charge
  densities from elastic electron scattering}}, \bibinfo{journal}{Nucl. Phys.
  A} \bibinfo{volume}{218} (\bibinfo{year}{1974}) \bibinfo{pages}{509--541}.

\bibitem[{Zhou(2020)}]{Zhou2020}
\bibinfo{author}{J.~Zhou}, \bibinfo{title}{{The sum-of-Gaussian
  parameterizations fitted with the available deuteron form factor data}},
  \urlprefix\url{https://github.com/TooLate0800/Deuteron_radius_fitting/tree/master/SOG_fitting},
  \bibinfo{year}{2020}.

\bibitem[{Gross(2020)}]{Gross:2019thk}
\bibinfo{author}{F.~Gross}, \bibinfo{title}{{Covariant Spectator Theory of $np$
  scattering: Deuteron form factors}}, \bibinfo{journal}{Phys. Rev. C}
  \bibinfo{volume}{101}~(\bibinfo{number}{2}) (\bibinfo{year}{2020})
  \bibinfo{pages}{024001}.

\bibitem[{Hummel and Tjon(1990)}]{Hummel:1990zz}
\bibinfo{author}{E.~Hummel}, \bibinfo{author}{J.~A. Tjon},
  \bibinfo{title}{{Relativistic analysis of meson exchange currents in elastic
  electron deuteron scattering}}, \bibinfo{journal}{Phys. Rev. C}
  \bibinfo{volume}{42} (\bibinfo{year}{1990}) \bibinfo{pages}{423--437}.

\bibitem[{Hummel and Tjon(1994)}]{Hummel:1993fq}
\bibinfo{author}{E.~Hummel}, \bibinfo{author}{J.~A. Tjon},
  \bibinfo{title}{{Relativistic description of electron scattering on the
  deuteron}}, \bibinfo{journal}{Phys. Rev. C} \bibinfo{volume}{49}
  (\bibinfo{year}{1994}) \bibinfo{pages}{21--39}.

\bibitem[{Higinbotham(2020)}]{Higinbotham2020}
\bibinfo{author}{D.~W. Higinbotham}, \bibinfo{title}{Parameterization of
  Deuteron Elastic Form Factors},
  \urlprefix\url{https://github.com/dhiginbotham/DeuteronFormFactors},
  \bibinfo{year}{2020}.

\bibitem[{Ding et~al.(2020)Ding, Raya, Binosi, Chang, Roberts, and
  Schmidt}]{Ding:2019qlr}
\bibinfo{author}{M.~Ding}, \bibinfo{author}{K.~Raya},
  \bibinfo{author}{D.~Binosi}, \bibinfo{author}{L.~Chang},
  \bibinfo{author}{C.~D. Roberts}, \bibinfo{author}{S.~M. Schmidt},
  \bibinfo{title}{{Drawing insights from pion parton distributions}},
  \bibinfo{journal}{Chin. Phys. C (Lett.)} \bibinfo{volume}{44}
  (\bibinfo{year}{2020}) \bibinfo{pages}{031002}.

\bibitem[{Eichmann et~al.(2022)Eichmann, Ferreira, and
  Stadler}]{Eichmann:2021vnj}
\bibinfo{author}{G.~Eichmann}, \bibinfo{author}{E.~Ferreira},
  \bibinfo{author}{A.~Stadler}, \bibinfo{title}{{Going to the light front with
  contour deformations}}, \bibinfo{journal}{Phys. Rev. D}
  \bibinfo{volume}{105}~(\bibinfo{number}{3}) (\bibinfo{year}{2022})
  \bibinfo{pages}{034009}.

\bibitem[{Yao et~al.(2022)Yao, Binosi, Cui, and Roberts}]{Yao:2021pdy}
\bibinfo{author}{Z.-Q. Yao}, \bibinfo{author}{D.~Binosi},
  \bibinfo{author}{Z.-F. Cui}, \bibinfo{author}{C.~D. Roberts},
  \bibinfo{title}{{Semileptonic transitions: $B_{(s)} \to \pi(K)$; $D_s \to K$;
  $D\to \pi, K$; and $K\to \pi$}}, \bibinfo{journal}{Phys. Lett. B}
  \bibinfo{volume}{824} (\bibinfo{year}{2022}) \bibinfo{pages}{136793}.

\bibitem[{Abrams et~al.(2022)}]{Abrams:2021xum}
\bibinfo{author}{D.~Abrams}, et~al., \bibinfo{title}{{Measurement of the
  Nucleon $F^n_2/F^p_2$ Structure Function Ratio by the Jefferson Lab MARATHON
  Tritium/Helium-3 Deep Inelastic Scattering Experiment}},
  \bibinfo{journal}{Phys. Rev. Lett.}
  \bibinfo{volume}{128}~(\bibinfo{number}{13}) (\bibinfo{year}{2022})
  \bibinfo{pages}{132003}.

\bibitem[{Cui et~al.(2022)Cui, Gao, Binosi, Chang, Roberts, and
  Schmidt}]{Cui:2021gzg}
\bibinfo{author}{Z.-F. Cui}, \bibinfo{author}{F.~Gao},
  \bibinfo{author}{D.~Binosi}, \bibinfo{author}{L.~Chang},
  \bibinfo{author}{C.~D. Roberts}, \bibinfo{author}{S.~M. Schmidt},
  \bibinfo{title}{{Valence quark ratio in the proton}}, \bibinfo{journal}{Chin.
  Phys. Lett. \emph{Express}} \bibinfo{volume}{39}~(\bibinfo{number}{04})
  (\bibinfo{year}{2022}) \bibinfo{pages}{041401}.

\end{thebibliography}

\end{document}